%% file: ms.tex
\documentclass{emulateapj}  
\pdfoutput=1
\usepackage{epstopdf}
\usepackage{natbib}
\usepackage{ifthen}

\newcommand{\ispreprint}{true}

\usepackage{amsmath}
\newcommand{\imacs}{IMACS}
\newcommand{\ldss}{LDSS3}

\newcommand{\rab}[2]{#1^{\mathrm{h}}#2^{\mathrm{m}}}
\newcommand{\decb}[2]{#1^{\circ}#2{\arcmin}}

\newcommand{\deca}[1]{#1^{\circ}}

\newcommand{\msun}{\ensuremath{M_\odot}}

\newcommand{\ngal}{\ensuremath{N_{\mathrm{gal}}}}
\newcommand{\mtwohundred}{\ensuremath{M_{200}}}
\newcommand{\ntwohundred}{\ensuremath{N_{200}}}
\newcommand{\rtwohundred}{\ensuremath{R_{200}}}

\newcommand{\sigmars}{\ensuremath{\sigma^{\mathrm{rs}}}}
\newcommand{\dd}{\ensuremath{\mathrm{d}}}
\newcommand{\specz}{\ensuremath{z_{\mathrm{spec}}}}

\newcommand{\redseqz}{\ensuremath{z_{\mathrm{rs}}}}
\newcommand{\uk}{\mbox{$\mu \mbox{K}$}}

\slugcomment{Submitted to \apj}

\shorttitle{Redshift and Richness Estimates for 21 SZ Clusters}
\shortauthors{High et al.}

\def\Harvard{1}
\def\Illinois{2}
\def\Cardiff{3}
\def\UChicago{4}
\def\Fermilab{5}
\def\NCSA{6}
\def\UNDakota{7}
\def\Berkeley{8}
\def\KICPChicago{9}
\def\EFIChicago{10}
\def\IAPFrance{11}
\def\LANL{12}
\def\PhysicsUChicago{13}
\def\CfA{14}
\def\AAUChicago{15}
\def\McGill{16}
\def\Colorado{17}
\def\UChile{18}
\def\UCSC{19}
\def\NASA{20}
\def\LBNL{21}
\def\UTokyo{22}
\def\Michigan{23}
\def\Munich{24}
\def\ExcellenceCluster{25}
\def\MPE{26}
\def\CaseWestern{27}
\def\Yale{28}
\def\NOAO{29}
\def\UPitt{30}

\begin{document}

\title{Optical Redshift and Richness Estimates for Galaxy Clusters\\
  Selected with the Sunyaev-Zel'dovich Effect from\\ 2008 South Pole
  Telescope Observations}

\author{
F.~W.~High,\altaffilmark{\Harvard}
B.~Stalder,\altaffilmark{\Harvard}
J.~Song,\altaffilmark{\Illinois}
P.~A.~R.~Ade,\altaffilmark{\Cardiff}
K.~A.~Aird,\altaffilmark{\UChicago}
S.~S.~Allam,\altaffilmark{\Fermilab} 
R.~Armstrong,\altaffilmark{\NCSA} 
W.~A.~Barkhouse,\altaffilmark{\UNDakota} 
B.~A.~Benson,\altaffilmark{\Berkeley,\KICPChicago,\EFIChicago}
E.~Bertin,\altaffilmark{\IAPFrance} 
S.~Bhattacharya,\altaffilmark{\LANL}
L.~E.~Bleem,\altaffilmark{\KICPChicago,\PhysicsUChicago}
M.~Brodwin,\altaffilmark{\CfA}
E.~J.~Buckley-Geer,\altaffilmark{\Fermilab} 
J.~E.~Carlstrom,\altaffilmark{\KICPChicago,\EFIChicago,\PhysicsUChicago,\AAUChicago}
P.~Challis,\altaffilmark{\CfA}
C.~L.~Chang,\altaffilmark{\KICPChicago,\EFIChicago}
T.~M.~Crawford,\altaffilmark{\KICPChicago,\AAUChicago}
A.~T.~Crites,\altaffilmark{\KICPChicago,\AAUChicago}
T.~de Haan,\altaffilmark{\McGill}
S.~Desai,\altaffilmark{\NCSA} 
M.~A.~Dobbs,\altaffilmark{\McGill}
J.~P.~Dudley,\altaffilmark{\McGill}
R.~J.~Foley,\altaffilmark{\CfA} 
E.~M.~George,\altaffilmark{\Berkeley}
M.~Gladders,\altaffilmark{\KICPChicago,\AAUChicago}
N.~W.~Halverson,\altaffilmark{\Colorado}
M.~Hamuy,\altaffilmark{\UChile}
S.~M.~Hansen,\altaffilmark{\UCSC} 
G.~P.~Holder,\altaffilmark{\McGill}
W.~L.~Holzapfel,\altaffilmark{\Berkeley}
J.~D.~Hrubes,\altaffilmark{\UChicago}
M.~Joy,\altaffilmark{\NASA}
R.~Keisler,\altaffilmark{\KICPChicago,\PhysicsUChicago}
A.~T.~Lee,\altaffilmark{\Berkeley,\LBNL}
E.~M.~Leitch,\altaffilmark{\KICPChicago,\AAUChicago}
H.~Lin,\altaffilmark{\Fermilab} 
Y.-T.~Lin,\altaffilmark{\UTokyo} 
A.~Loehr,\altaffilmark{\CfA}
M.~Lueker,\altaffilmark{\Berkeley}
D.~Marrone,\altaffilmark{\UChicago,\KICPChicago}
J.~J.~McMahon,\altaffilmark{\KICPChicago,\EFIChicago,\Michigan}
J.~Mehl,\altaffilmark{\KICPChicago,\AAUChicago}
S.~S.~Meyer,\altaffilmark{\KICPChicago,\EFIChicago,\PhysicsUChicago,\AAUChicago}
J.~J.~Mohr,\altaffilmark{\Munich,\ExcellenceCluster,\MPE}
T.~E.~Montroy,\altaffilmark{\CaseWestern}
N.~Morell,\altaffilmark{\UChile}
C.-C.~Ngeow,\altaffilmark{\Illinois} 
S.~Padin,\altaffilmark{\KICPChicago,\AAUChicago}
T.~Plagge,\altaffilmark{\Berkeley,\AAUChicago}
C.~Pryke,\altaffilmark{\KICPChicago,\EFIChicago,\AAUChicago}
C.~L.~Reichardt,\altaffilmark{\Berkeley}
A.~Rest,\altaffilmark{\Harvard} 
J.~Ruel,\altaffilmark{\Harvard} 
J.~E.~Ruhl,\altaffilmark{\CaseWestern}
K.~K.~Schaffer,\altaffilmark{\KICPChicago,\EFIChicago}
L.~Shaw,\altaffilmark{\McGill,\Yale} 
E.~Shirokoff,\altaffilmark{\Berkeley}
R.~C.~Smith,\altaffilmark{\NOAO} 
H.~G.~Spieler,\altaffilmark{\LBNL}
Z.~Staniszewski,\altaffilmark{\CaseWestern}
A.~A.~Stark,\altaffilmark{\CfA}
C.~W.~Stubbs,\altaffilmark{\Harvard,\CfA}
D.~L.~Tucker,\altaffilmark{\Fermilab} 
K.~Vanderlinde,\altaffilmark{\McGill}
J.~D.~Vieira,\altaffilmark{\KICPChicago,\PhysicsUChicago}
R.~Williamson,\altaffilmark{\KICPChicago,\AAUChicago}
W.~M.~Wood-Vasey,\altaffilmark{\UPitt}
Y.~Yang,\altaffilmark{\Illinois}
O.~Zahn,\altaffilmark{\Berkeley} and
A.~Zenteno\altaffilmark{\Munich,\ExcellenceCluster}
}

\altaffiltext{\Harvard}{Department of Physics, Harvard University, 17 Oxford Street, Cambridge, MA 02138}
\altaffiltext{\Illinois}{Department of Astronomy, University of Illinois, 1002 West Green Street, Urbana, IL 61801}
\altaffiltext{\Cardiff}{Department of Physics and Astronomy, Cardiff University, CF24 3YB, UK}
\altaffiltext{\UChicago}{University of Chicago, 5640 South Ellis Avenue, Chicago, IL 60637}
\altaffiltext{\Fermilab}{Fermi National Accelerator Laboratory, P.O.\ Box 500, Batavia, IL 60510}
\altaffiltext{\NCSA}{National Center for Supercomputing Applications, University of Illinois, 1205 West Clark Street, Urbana, IL 61801}
\altaffiltext{\UNDakota}{Department of Physics \& Astrophysics, University of North Dakota, Grand Forks, ND 58202}
\altaffiltext{\Berkeley}{Department of Physics, University of California, Berkeley, CA 94720}
\altaffiltext{\KICPChicago}{Kavli Institute for Cosmological Physics, University of Chicago, 5640 South Ellis Avenue, Chicago, IL 60637}
\altaffiltext{\EFIChicago}{Enrico Fermi Institute, University of Chicago, 5640 South Ellis Avenue, Chicago, IL 60637}
\altaffiltext{\IAPFrance}{Institut d’Astrophysique de Paris, 98bis bd Arago, F-75014 Paris, France}
\altaffiltext{\LANL}{Los Alamos National Laboratory, Los Alamos, New Mexico}
\altaffiltext{\PhysicsUChicago}{Department of Physics, University of Chicago, 5640 South Ellis Avenue, Chicago, IL 60637}
\altaffiltext{\CfA}{Harvard-Smithsonian Center for Astrophysics, 60 Garden Street, Cambridge, MA 02138}
\altaffiltext{\AAUChicago}{Department of Astronomy and Astrophysics, University of Chicago, 5640 South Ellis Avenue, Chicago, IL 60637}
\altaffiltext{\McGill}{Department of Physics, McGill University, 3600 Rue University, Montreal, Quebec H3A 2T8, Canada}
\altaffiltext{\Colorado}{Department of Astrophysical and Planetary Sciences and Department of Physics, University of Colorado, Boulder, CO 80309}
\altaffiltext{\UChile}{Departamento de Astronomia, Universidad de Chile, Casilla 36-D, Santiago, Chile} 
\altaffiltext{\UCSC}{University of California Observatories \& Department of Astronomy, University of California, Santa Cruz, CA 95064}
\altaffiltext{\NASA}{Department of Space Science, VP62, NASA Marshall Space Flight Center, Huntsville, AL 35812}
\altaffiltext{\LBNL}{Physics Division, Lawrence Berkeley National Laboratory, Berkeley, CA 94720}
\altaffiltext{\UTokyo}{(Institute for Physics and Mathematics of the Universe, University of Tokyo, 5-1-5 Kashiwa-no-ha, Kashiwa-shi, Chiba 277-8568, Japan}
\altaffiltext{\Michigan}{Department of Physics, University of Michigan, 450 Church Street, Ann Arbor, MI, 48109}
\altaffiltext{\Munich}{Department of Physics, Ludwig-Maximilians-Universit\"{a}t, Scheinerstr.\ 1, 81679 M\"{u}nchen, Germany}
\altaffiltext{\ExcellenceCluster}{Excellence Cluster Universe, Boltzmannstr.\ 2, 85748 Garching, Germany}
\altaffiltext{\MPE}{Max-Planck-Institut f\"{u}r extraterrestrische Physik, Giessenbachstr.\ 85748 Garching, Germany}
\altaffiltext{\CaseWestern}{Physics Department, Case Western Reserve University, Cleveland, OH 44106}
\altaffiltext{\Yale}{Department of Physics, Yale University, P.O.\ Box 208210, New Haven, CT 06520-8120}
\altaffiltext{\NOAO}{Cerro Tololo Interamerican Observatory, La Serena, Chile}
\altaffiltext{\UPitt}{University of Pittsburgh, 3941 O'Hara St., Pittsburgh, PA 15260}




\email{high@physics.harvard.edu}

\begin{abstract}

  We present redshifts and optical richness properties of 21 galaxy clusters
  uniformly selected by their Sunyaev-Zel'dovich signature.
  These clusters, plus an additional, unconfirmed candidate, were detected 
  in a $178\,\deg^2$ 
  area surveyed by the South Pole Telescope in 2008.  Using $griz$
  imaging from the
  Blanco Cosmology Survey and from pointed Magellan telescope
  observations, as well as spectroscopy using Magellan facilities, we confirm the existence
  of clustered red-sequence galaxies, report red-sequence photometric redshifts,
  present spectroscopic redshifts for a subsample, and derive 
  \rtwohundred\ radii and \mtwohundred\ masses from optical
  richness.  The clusters span redshifts from 0.15 to greater than 1, with a
  median redshift of 0.74; three clusters are estimated to be at $z > 1$. Redshifts
  inferred from mean red-sequence colors exhibit 2\% 
  RMS scatter in $\sigma_z/(1+z)$ with respect to the spectroscopic
  subsample for $z < 1$.  We show that \mtwohundred\ cluster masses
  derived from optical richness correlate with masses
  derived from South Pole Telescope data and agree with previously
  derived scaling relations to within the uncertainties.  
  Optical and infrared
  imaging is an efficient means of cluster identification and
  redshift estimation in large Sunyaev-Zel'dovich surveys, and exploiting the
  same data for richness measurements, as we have done, will be useful
  for constraining cluster masses and radii for large samples in cosmological
  analysis. 

\end{abstract}


\defcitealias{vanderlinde10}{V10}
\defcitealias{staniszewski08}{S09}
\defcitealias{menanteau08}{M09}
\defcitealias{menanteau09}{MH09}
\defcitealias{menanteau10}{M10}
\defcitealias{abell89}{A89}

\newpage


\section{Introduction}

Galaxy clusters are laboratories for both astrophysics and cosmology
\citep{evrard04}.  
Clusters represent the most massive dark matter halos, and their 
number density as a function of cosmic time
is highly sensitive to dark energy
\citep{wang98,haiman01,holder01b,battye03,molnar04,wang04,lima07}.
The mass of these systems is dominated by dark matter, 
but the primary means of observing clusters---especially large samples
of them---are the luminous baryons of the hot intracluster gas and the
galaxies themselves.  The formation of the halos is well understood,
while the precise behavior of the baryons is not as well modeled
\citep[see][for a review]{voit04}.  This gap must be closed so that
data from large cluster surveys can place precise constraints on cosmological
parameters over a wide range of redshifts.  Multi-wavelength observations
of a cleanly selected, redshift-independent sample of galaxy clusters
are a potentially powerful method of achieving this.

Searches for galaxy clusters using the Sunyaev-Zel'dovich (SZ) effect \citep[][]{sunyaev72}
promise to provide such a clean, redshift-independent sample.  The SZ effect is
scattering of cosmic microwave background photons to higher energy by
the hot electrons in galaxy clusters \citep{birkinshaw99}.  The SZ
surface brightness is independent of
redshift but is closely related to cluster mass and so it is
expected to be an excellent method for creating 
approximately mass-limited samples extending over a wide redshift range
\citep[][]{carlstrom02}.  The constraints on
cosmological parameters from such samples are complementary to
geometrical tests using type Ia supernovae and baryon acoustic
oscillations \citep[e.g.,][]{vikhlinin09}.  Two SZ surveys, the
Atacama Cosmology Telescope \citep[ACT;][]{fowler07} and the South
Pole Telescope \citep[SPT;][]{carlstrom09} projects, are well
positioned to provide large surveys which can be used for growth of
structure studies.



\citet[hereafter S09]{staniszewski08} presented the first
discovery of previous unknown galaxy clusters using 
their SZ signature.
Cluster redshifts are needed in addition to
SZ data to provide the strongest constraints on dark energy.
Coordinated optical follow-up observations can provide the needed
redshift measurements.  
The Blanco Cosmology Survey \citep[BCS;][and
\url{http://cosmology.illinois.edu/BCS/}]{ngeow06}, an NOAO survey
program (2005-2008), provided multiband optical observations for the
initial follow-up of portions of the first SPT survey fields.
These data were used to identify optical counterparts to the
\citetalias{staniszewski08} sample, search for giant arcs, explore possible cluster
superpositions, and derive photometric redshifts.

Cluster 
mass can
be estimated using several methods: the SZ and X-ray luminosity, which are
sensitive to intracluster electrons; the number, 
luminosity, and velocity dispersion of cluster galaxies; and from
gravitational lensing, which is the most direct probe
of total cluster mass.  
\citet{menanteau09}
characterized the galaxy 
counts and luminosity of the \citetalias{staniszewski08} cluster sample, and
\citet{mcinnes09} subsequently explored their weak gravitational
lensing signals.

Using data acquired by the SPT in 2008, \citet[][hereafter
V10]{vanderlinde10} present an additional 17 SZ-detected clusters.
Here we describe coordinated optical imaging of the catalog of 21
uniformly selected SZ detections, and new spectroscopic results on 8 of the
clusters.  Counterparts to a subset have been found
in the catalogs of \citet[][hereafter A89]{abell89} and
\citet[][hereafter SCS-II]{menanteau10}. 
Seven clusters fell within the BCS footprint.  
 For
the 
remaining 14 clusters, and also for a subset of the BCS sample, we
conducted pointed imaging observations  
and, for 8 clusters, spectroscopic observations,
with the Magellan telescopes.  The photometry was used to
search for overdensities of red-sequence galaxies near the SZ
locations, and if present, estimate their redshifts and also characterize
their mass via optical red-sequence galaxy counts, or {\it richness}.

We describe in \S \ref{sec:data} the observations and data reduction.
Section \ref{sec:analysis} outlines the redshift and richness analysis
we used, and \S \ref{sec:results} describes the results on redshift
(\S \ref{sec:redshifts}) and richness (\S \ref{sec:richness}).  In
Section \ref{sec:discussion} we discuss the results, and conclude
with \S \ref{sec:conclusion}.

Throughout this paper we assume a flat concordance $\Lambda\mathrm{CDM}$
universe, with $(\Omega_{\Lambda},\Omega_{\mathrm{M}},h) =
(0.736,0.264,0.71)$ \citep{dunkley09}.  All magnitudes are in the Sloan
Digital Sky Survey (SDSS) $griz$ AB system.


\section{Data Acquisition \& Reduction}
\label{sec:data}

Cluster detection was achieved using millimeter-wavelength data from the South Pole
Telescope, and optical imaging and spectroscopy provided cluster
confirmation and redshift and richness estimates.

\subsection{South Pole Telescope}
\label{sec:spt}

The sample of 21 clusters presented here and in \citetalias{vanderlinde10}
is the first cosmologically significant catalog of clusters selected via the SZ effect.
The sample was selected from two SPT survey fields totaling
$178\deg^2$ at
$\mathrm{R.A.}=\rab{23}{30}$, $\mathrm{Decl.}=\deca{-55}$ and
$\mathrm{R.A.}=\rab{5}{30}$, $\mathrm{Decl.}=\deca{-55}$ (J2000).
Both fields were observed with arcminute resolution
to an equivalent white noise level of 18 $\uk$-arcmin\footnote{The unit $\textrm{K}$
refers to equivalent fluctuations in the CMB temperature, i.e., the
  level of temperature fluctuation of a 2.73$\,$K blackbody that would
  be required to produce the same power fluctuation. See
  \citetalias{vanderlinde10}.}. 

The SPT time-ordered data were filtered and binned into maps, with the filtering
acting roughly as a $1^\circ$ high-pass filter in the R.A.\ direction. Clusters
were extracted from these maps using a matched filter approach based on the work
of \citep{haehnelt96,herranz02a,herranz02b,melin06}. Spatial filters were constructed to maximize detection
significance within a set of cluster profiles.  
The SPT astrometry is based on comparisons of radio source positions
derived from SPT maps and positions of those sources in the AT20G
catalog \citep{murphy10}, and should be accurate to $5\arcsec$.

Cluster candidates were then identified by selecting all
peaks above a fixed significance threshold and choosing the filter scale which
produced the maximum detection significance $\xi$.
A 3 parameter model for $M_{200}$ involving $\xi$ and the cluster redshift
is presented in \citetalias{vanderlinde10}, along with the details of
the SPT data reduction.

\subsection{Blanco Cosmology Survey}
\label{sec:bcs}

BCS is an NOAO survey program to obtain deep $griz$
imaging of two southern fields centered at
$\mathrm{R.A.}=\rab{23}{00}$, $\mathrm{Decl.}=\decb{-55}{12}$ and
$\mathrm{R.A.}=\rab{5}{30}$, $\mathrm{Decl.}=\decb{-52}{47}$ (J2000), each
roughly $50\,\deg^2$.  The 2008 SPT survey fields are larger than 
the BCS fields and include the entire BCS regions.  BCS was conducted
from 2005--2008 using the Mosaic-II wide-field imager on the Blanco
4~m telescope at Cerro Tololo Inter-American Observatory. Mosaic-II is
an array of eight $2\,\mathrm{k}\times 4\,\mathrm{k}$ CCDs with a
pixel scale of $0.270\arcsec$~pixel$^{-1}$ and a $0.36\,\deg^2$ field
of view. The strategy was to obtain deep, contiguous $griz$ imaging of
the survey fields.  In addition, BCS
imaging was carried out for 7 photometric redshift
calibration fields which include a sample of several
thousand published spectroscopic redshifts.

The BCS data were processed and
calibrated using a data management system developed for the
Dark Energy Survey \citep{ngeow06,mohr08} and run on the NCSA TeraGrid IA-64 Linux cluster. Data
reduction includes crosstalk correction, overscan correction, bias
subtraction, flat fielding, fringe and illumination corrections, field
distortion correction, standard star photometric calibration,
coadd-stacking, and photometric extraction of sources. 
BCS stacks typically reach $5\sigma$ galaxy photometry limits of 24.75,
24.65, 24.35 and 23.5 mag in $griz$, corresponding, for example, to a
$0.5L^*$ cluster elliptical at $z=1$.  Completeness measurements in a
typical field that are derived from comparison with deeper, better
seeing CFHT data suggest 50\%
completeness limits of 24.25, 24.0, 23.75 and 23.0 in $griz$ (Zenteno
et al.\ in prep.).

\subsection{Magellan}
\label{sec:magellan}

\subsubsection{Imaging}

For clusters that fell outside the BCS
coverage region, for 5 that were within the BCS region, and for the unconfirmed candidate (see \S \ref{sec:clusters})
we obtained $griz$ imaging with the Inamori
Magellan Areal Camera and Spectrograph \citep[IMACS;][]{dressler03,
osip08} and Low Dispersion Survey Spectrograph \citep[LDSS3;
see][]{osip08}---both in imaging mode---on the twin Magellan
$6.5\,\mathrm{m}$ telescopes.  
IMACS is on the Magellan Baade 
telescope at the $f/11$ Nasmyth focus.  Its circular field of view in
$f/2$ mode subtends $0.20\,\deg^2$, mapped onto a $8192 \times 8192$
pixel, $8$-chip CCD array, for a pixel scale of
$0.200\arcsec$~pixel$^{-1}$. LDSS3 is on the Magellan Clay telescope
at the $f/11$ Nasmyth focus, and its roughly $60\,\mathrm{arcmin}^2$
circular field of view maps onto a subregion of a $4064 \times 4064$
pixel CCD, at a pixel scale of $0.189\arcsec$~pixel$^{-1}$.

Using the science-tested pipeline described by
\citet{rest05a} and \citet{miknaitis07}, the same image reduction
operations described above were performed on the Magellan images.
Instead of using standard stars for photometric calibration, however,
we employ Stellar Locus Regression \citep[SLR;][]{high09} using stars
that appear in the cluster images themselves.  SLR delivers
photometric calibrations by regressing the instrumental color-color
locus of stars in any field to the known, astrophysically fundamental
locus in the AB system.  SLR enabled us to forego observations of
standard stars altogether, maximizing the total integration time on
the cluster fields.

The principal difference in strategy between Magellan and BCS
observations is that the Magellan exposure times were adaptive rather
than uniform.  We first exposed for roughly $100\,\mathrm{s}$ in
$griz$, searched for a cluster in the images, and continued with
additional exposures if none was found. In median seeing of
$0.8\arcsec$, these exposures reach nominal,
$5\sigma$ point-source limiting magnitudes of 24.8, 24.8, 24.4, and
23.4~mag in $griz$.
If no cluster was detected with the initial set of images, we acquired
further exposures until a detection was achieved at a depth of
approximately $1$ to $0.4L^*$ with respect to the early-type cluster galaxies.
This strategy results in highly variable depth for the Magellan imaging, but is the
most efficient use of telescope time for follow-up observations.

\ifthenelse{\equal{\ispreprint}{true}}
{
  Combined optical and SZ images are shown in Appendix \ref{app:thumbs},
  Figures~\ref{fig:thumb1}--\ref{fig:thumb22}.

}
{
  \input{thumbs_inline}

}

\subsubsection{Spectroscopy}

Spectroscopic data were acquired with LDSS3 in long-slit mode for the
purpose of measuring redshifts.  Given limited telescope time for
spectroscopy, we observed only a subset of the confirmed clusters.
The subset was chosen to span the widest possible range in redshifts,
so that we could assess the performance of our red-sequence redshift
measurement methodology (see \S \ref{sec:rsredshifts} and \S
\ref{sec:redshifts}).

We obtained low-resolution spectra of galaxies in the field of
8 SPT clusters, using the VPH-Red and VPH-All grisms. The median
seeing was about $0.7\arcsec$ and conditions were photometric.
Standard CCD processing and 2D-spectrum extraction, with preliminary
wavelength solutions, were accomplished with the COSMOS reduction
package\footnote{http://obs.carnegiescience.edu/Code/cosmos/ }; the
1D spectra were then extracted using the {\sc apall} task in IRAF. We
employed our own IDL routines to flux calibrate the data and remove
telluric absorption using the well-exposed continua of the
spectrophotometric standards \citep{Wade88, Foley03}.


\section{Analysis}
\label{sec:analysis}

We adopted a standard red-sequence model for optical cluster
detection, photometric redshifts, and richness
estimation. As previously introduced, spectroscopic data on about half
the clusters were
used to empirically correct the red-sequence model colors
and then verify photometric red-sequence redshifts over a long
redshift baseline.

\subsection{Red-Sequence Model}
\label{sec:rs}

Red-sequence models were derived from the work of \citet{bruzual03},
reflecting passively evolving, instantaneous-burst stellar populations
with a formation redshift of $z = 3$, using \citet{bertelli94}
evolutionary tracks and the \citet{chabrier03} initial mass
function. At each redshift, a range of metallicities was chosen by
including a randomization in the metallicity-luminosity relation. The
models were smoothed by linear fits in color-magnitude space at each
redshift, and finally interpolated to arbitrary redshift using cubic
splines.

\begin{figure}
 \epsscale{1.25}
  \plotone{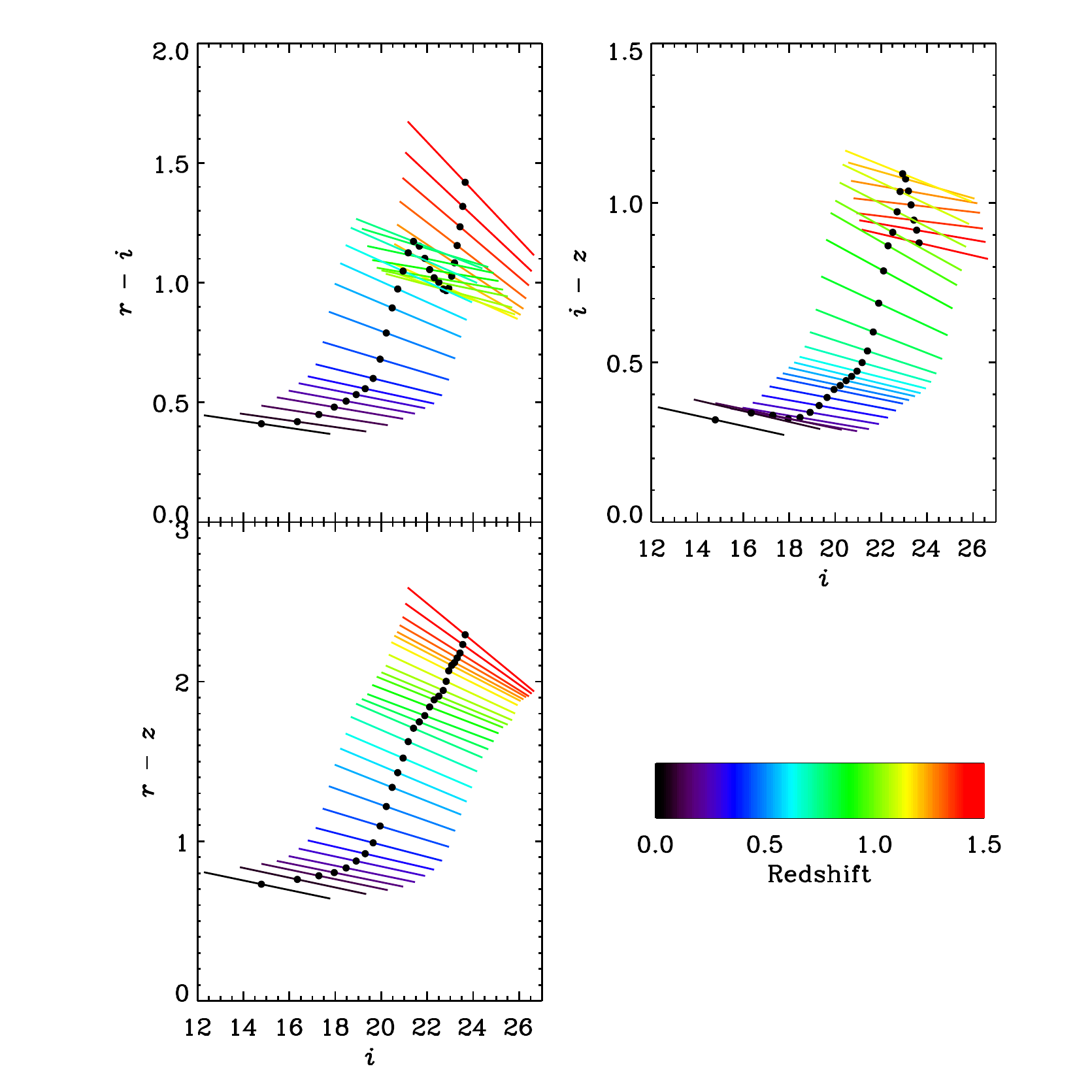}
 \caption{Red-sequence color-magnitude models as a function of
    redshift. Galaxy colors $r-i$ and $i-z$ were fit simultaneously in
    the red-sequence analysis, because that particular combination of
    colors (namely, $r-z = (r-i)+(i-z)$) allows for monotonic mapping
    from color to redshift over the widest redshift range ($0 < z
    \lesssim 1.4$) using optical wavelengths. The apparent $i$ band
    magnitude of $M^*$ at each redshift is denoted with the black
    points.\label{fig:rs}}
\end{figure}

We made an empirical correction to the model using the subsample of 10
clusters with spectroscopic data.  After performing the red sequence
peak finding (\S \ref{sec:rsredshifts}), we plotted red sequence
redshifts against spectroscopic redshifts.  The best fit line through
the data was measured to have slope $0.89\pm0.03$ and $y$-intercept
$0.04\pm0.02$.  Model colors were then corrected by reassigning the
model redshifts to be equal to the inverse of this linear relation.
The effect of the model color correction is to leave redshifts near
the pivot, $z\sim0.4$, roughly unchanged, and to boost redshifts near
$\sim1$ by about $\Delta z =+0.07$, or $\Delta z = +3.5\%(1+z)$. The
resulting red-sequence model is shown in Figure~\ref{fig:rs}.

\subsection{Red-Sequence Finding and Redshifts}
\label{sec:rsredshifts}

We confirm the existence of clusters by searching near the SZ cluster coordinate for a
background-subtracted excess of red-sequence objects, effectively segregated by
redshift. At each redshift step of
size $\Delta z = 0.01$ in the range $0.1<z<1.4$, and with an aperture
of $2\arcmin$ radius around the SZ coordinate, we select all objects
with a photometric signal-to-noise ratio $>5$ in $riz$, whose
$r-i$ and $i-z$ colors are also within $2\sigma$
of the red-sequence model line. The total uncertainty, e.g., for $r-i$,
is defined as
\begin{equation}
  \sigma^{2} \equiv \sigma_{ri}^2 + {\sigmars_{ri}}^2,
\end{equation}
and likewise for $i-z$. Here $\sigma_{ri}$ ($\sigma_{iz}$) is the
photometric uncertainty in $r-i$ ($i-z$) color of an object, and
$\sigmars_{ri}$ ($\sigmars_{iz}$) is the intrinsic color scatter of
the red sequence. These colors were chosen because their combination,
$r-z = (r-i)+(i-z)$, increases monotonically over a long baseline of
redshift, $0 < z \lesssim 1.4$.  We assume
$(\sigmars_{ri},\sigmars_{iz}) = (0.05,0.05)\,\mathrm{mag}$
\citep{koester07,menci08,mei09}.  The intrinsic scatter in color of
cluster ellipticals alone is about two times smaller than this, and
has been shown to be constant with redshift out to $z\approx 1.2$
\citep{menci08,mei09}.  Morphological and spectral galaxy
classification is beyond the scope of this work, and we assume our red
sequence selections contain S0 galaxies in addition to ellipticals.
This increases the color scatter and may also effectively induce
redshift evolution due to evolving S0 populations; we ignore the
latter effect, as the small number of clusters presented here is not
sufficient to constrain redshift evolution.

We sum the selected galaxies, and normalize the counts by the projected
area. This yields total surface density of all objects with colors
consistent with the red-sequence model, within the $2\arcmin$
aperture, as a function of redshift,
$\Sigma_{\mathrm{total}}(z)$. Uncertainties are estimated as the
square root of counts, divided by the area.

The background red-sequence surface density is measured in a similar
way.  In the same set of $riz$ exposures, we count red-sequence
objects within many adjacent apertures of size
$5\arcmin\times5\arcmin$ over the entire field of view, excluding the
$2\arcmin$ region around the SPT candidate position. The background surface
density at each redshift, $\Sigma_{\mathrm{background}}(z)$, is
calculated as the median of area-normalized counts from all apertures,
and the uncertainty is the standard deviation divided by the square
root of the total
number of apertures. This approach is meant to minimize the
contamination of the background signal by the red galaxies in the
clusters themselves. 

The red-sequence excess as a function of redshift is 
\begin{equation}
  \label{eqn:rssigma}
  \Sigma_{\mathrm{net}} = \Sigma_{\mathrm{total}} - \Sigma_{\mathrm{background}}
\end{equation}
in units of galaxies per arcmin$^{2}$.  We then renormalize the red
sequence excess $\Sigma_{\mathrm{net}}$ in each redshift bin such that
each galaxy contributes a {\it constant area} under the curve.  The
reason for this is that one galaxy may fall into multiple adjacent
bins due to the size of color uncertainties, sometimes significantly
widening the red-sequence peaks, especially for faint objects.  The
renormalization is a measure to mitigate this effect.

A cluster is detected if (1) there is an excess of red galaxies of the
same apparent color in false color images, and (2) the galaxies
corresponding to the maximum overdensity,
$\Sigma_{\mathrm{net}}(z_{\mathrm{max}})$, are those identifiable in
the false-color images. The cluster photometric redshift is taken to be 
$z_{\mathrm{max}}$. Figure~\ref{fig:zhist} illustrates the process of
red-sequence finding for one of the SZ clusters.

\ifthenelse{\equal{\ispreprint}{true}}
{
\begin{figure*}
}
{
\begin{figure}
}
 \plotone{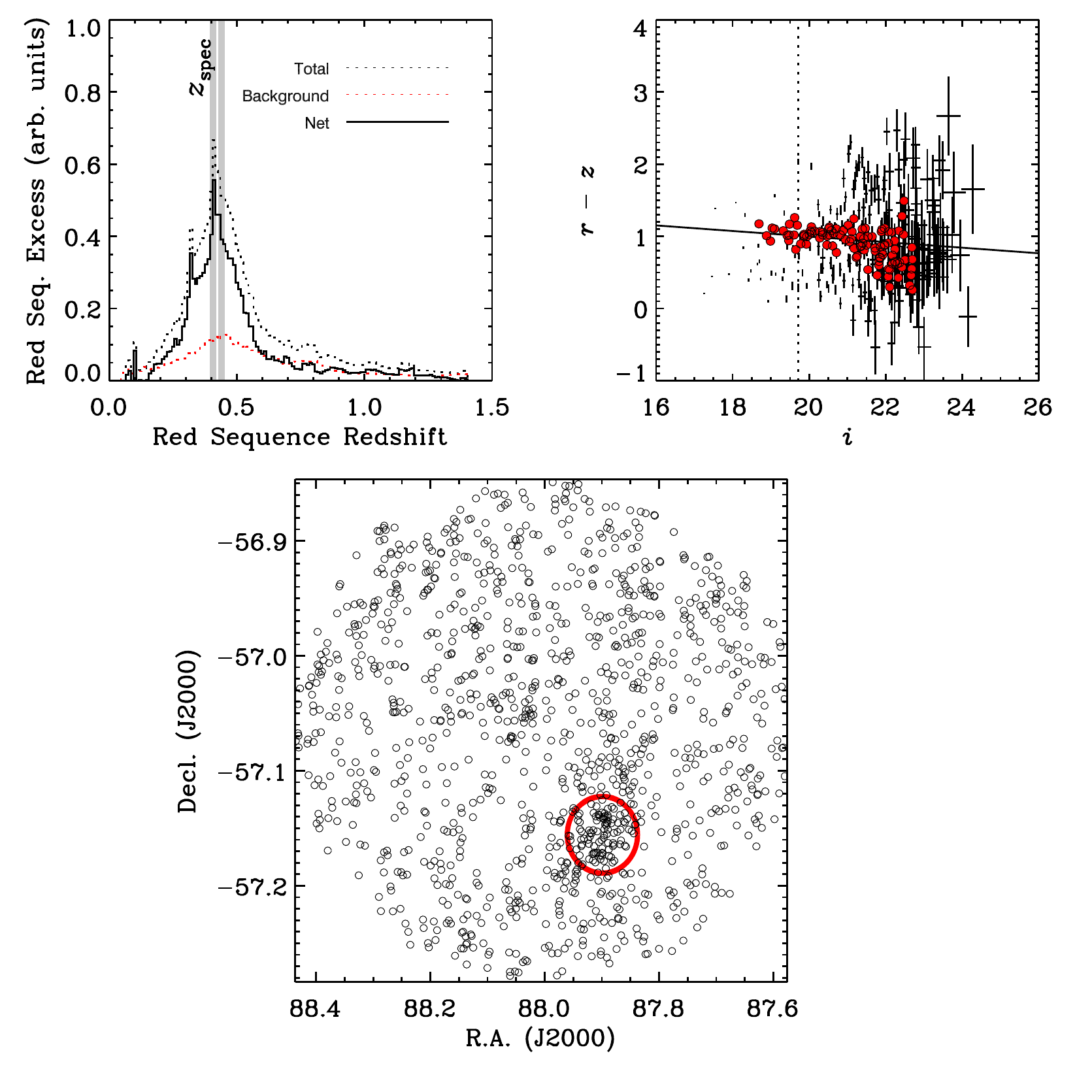}
\caption{An illustration of red-sequence finding for the cluster
   SPT-CL J0551-5709 at $\specz=0.4230$. The
   top left plot shows the surface density of objects consistent
   within $2\sigma$ of
   the red-sequence model as a function of redshift. The peak occurs
   at redshift $z_{\mathrm{max}} =0.41$, which is consistent with the
   spectroscopic redshift (vertical white line, with
   $\sigma_z=2\%(1+z)$ uncertainty region shaded in gray for
   illustration). The color-magnitude diagram for all objects within
   $2\arcmin$ of the SPT coordinate is shown in the upper right, and
   the subsample of red-sequence objects at $z_{\mathrm{max}}$ shown
   as red points.  The vertical dotted line is the model $m^*$ at this
   redshift. Positions of all objects consistent with the
   $z_{\mathrm{max}}$ red sequence are shown in the bottom panel,
   where we have also circled the cluster aperture. A spatial
   overdensity of objects is clearly
   seen at the aperture.\label{fig:zhist}}
\ifthenelse{\equal{\ispreprint}{true}}
{
\end{figure*}
}
{
\end{figure}
}


\subsubsection{Completeness}
\label{sec:completeness}

We tested for completeness in red-sequence finding at representative
BCS and Magellan depths.  After constructing mock optical catalogs
(Song et al.\ in prep.) containing simulated clusters of mass above
$\sim 3\times10^{14}\,h^{-1}\,\msun$, we searched for red-sequence
overdensities as we have described.  Our estimated completeness is
the fraction of clusters recovered from the mock catalog.
These mock catalogs include clusters with red and blue galaxy
populations that are tuned to match the populations observed in real
clusters \citep[e.g.,][]{lin04a}.  The galaxy distribution in space is determined
using subhalo positions within high-resolution $N$-body simulations, and
so the mock cluster spatial, kinematic and color signatures are a good
match to those seen in real clusters.

\begin{figure}
 \epsscale{1.2}
\plotone{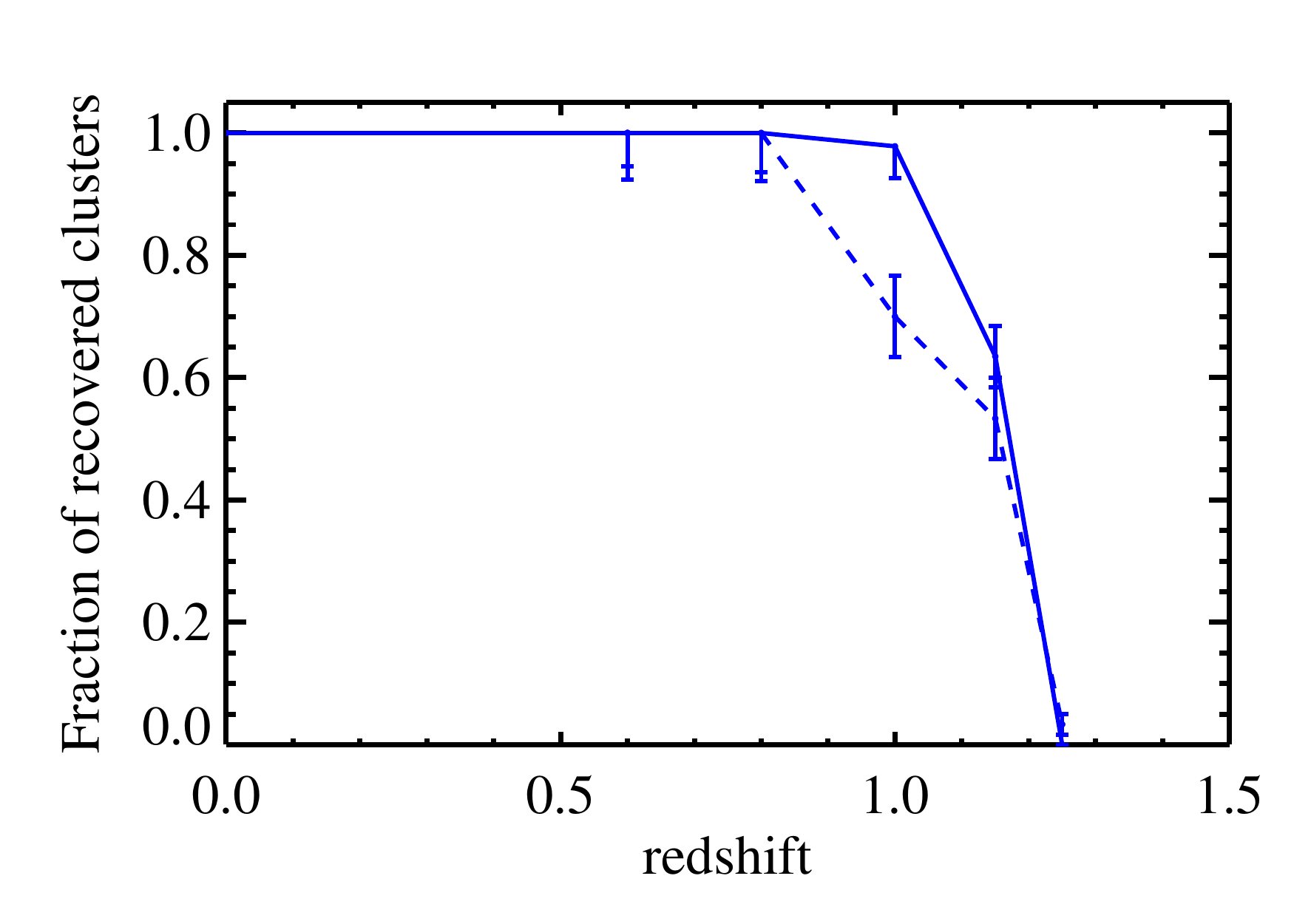}
\caption{This shows the completeness of optical cluster finding from
  tests on a mock galaxy catalog with depth representative of the BCS
  survey (solid line) and Magellan imaging (dashed line).  
  We have approximated Magellan data here as having limiting
  magnitudes $\sim1$ magnitude brighter than
  BCS.\label{fig:completeness}} 
\end{figure}

The resulting selection function is shown in Figure
\ref{fig:completeness}.  We recovered 100\% of simulated clusters above the given
mass threshold up to redshift $0.9$.  
At this point the $4000\,\mathrm{\AA}$ break begins to redshift out of the $i$ band,
making galaxies much harder to detect in $i-z$ color space.  The
completeness begins to fall here, and for these BCS and Magellan depths
the probability of finding an optical counterpart above this mass
threshold falls to zero by $z\approx1.2$.  Near-infrared photometry or much
deeper space based optical photometry is required to push reliably to
higher redshifts.  But for the current sample only a single SPT
cluster candidate was not confirmed using this multiband optical
method.

\subsection{$\ngal$ and $\ntwohundred$ Richness}
\label{sec:richnessanalysis}

After confirming a cluster and estimating its redshift, we measured
the optical richness using a procedure that emulates the MaxBCG
richness estimator \citep{koester07}, but adapted to high-redshift
clusters.  This began by again selecting objects with color within
$2\sigma$ of the red-sequence line at redshift $z_{\mathrm{max}}$;
luminosity brighter than $0.4L^{*}$ and fainter than the brightest
cluster galaxy (BCG); and
position within a projected radius of $R = 1\,h^{-1}\,\mathrm{Mpc}$ of the cluster
center.  We binned the selected objects in $i$ band magnitude bins of
size $\Delta m = 0.4$. We subtracted the background, as before, by
performing the same procedure in many apertures on the sky away from
the cluster but in the same exposures, and normalizing by projected
area. The background was subtracted from the total red-sequence counts
in each magnitude bin.

We then fitted Schechter luminosity functions
\citep[LF,][]{schechter76} to the $i$ band magnitudes of the selected
objects:
\begin{align}
  \phi(m)\dd m &= 0.4\ln(10) \phi^* 10^{-0.4(m-m^*)(\alpha+1)} \notag \\
               &\times {\mathrm{exp}}\left[{-10^{-0.4(m-m^*)}}\right]\dd m, \label{eqn:lf}
\end{align}
where $m^*$ is the characteristic magnitude of 
the LF, $\alpha$ is the faint-end slope, and
$\phi^*$ is the normalization.  Our photometry was not uniformly
complete on all clusters with respect to $m^*$, so our ability to
constrain the slope was weak.  We therefore fixed $\alpha=-1$,
which has been shown to be reasonable for the most massive MaxBCG clusters
\citep[][]{lin03a,hansen05,rudnik09,crawford09}.  We tested fixing $m^*$ at each
redshift using our passive model, and leaving it free.  For those
clusters for which the luminosity function was well constrained, we
measured values consistent with the model, and we ultimately chose to
fix this parameter at each redshift.  This is in agreement with the
detailed LF studies of Zenteno et al.\ (in prep) on a subset of these
clusters.  The faint-end slope $\alpha$ has 
been shown to evolve with redshift \citep{rudnik09}, however, such
studies must be performed at magnitudes fainter than $m^*+1$.  We test
varying $\alpha$ by $\pm 0.3$, and find our richness results are
largely unaffected to within our uncertainties.  
Because we integrate only to $m^*+1$ there is only weak sensitivity to
the adopted faint-end slope.

We fit the LF only at magnitudes brighter than our limiting
magnitude and fainter than the BCG.
Limiting magnitudes are estimated to be the faintest magnitude bin
before which the red-sequence background becomes incomplete, as
indicated by a deviation from linearity in the logarithmic $i$ band
magnitude distribution of red-sequence galaxies. 
We also assess 
photometric completeness using simulated point sources in our cluster
images.  We determine that we recover $90\%$ of simulated objects in
images at {\it all magnitudes} brighter than the above-defined limiting
magnitude. The majority of unrecovered objects are lost to pixel
masking due to bright stars and to object crowding.  We make a generic
correction by this amount in all magnitude bins, the effect of which
is to increase $\ngal$ estimates by $0.9^{-1}$.  The analytic LF is then
integrated down to $m^*+1$.  The resulting integration is an
estimate of $\ngal$, which is the number of red sequence galaxies
within $1\,h^{-1}\,\mathrm{Mpc}$ of the cluster center, above a fixed
luminosity threshold.

\begin{figure}
  \plotone{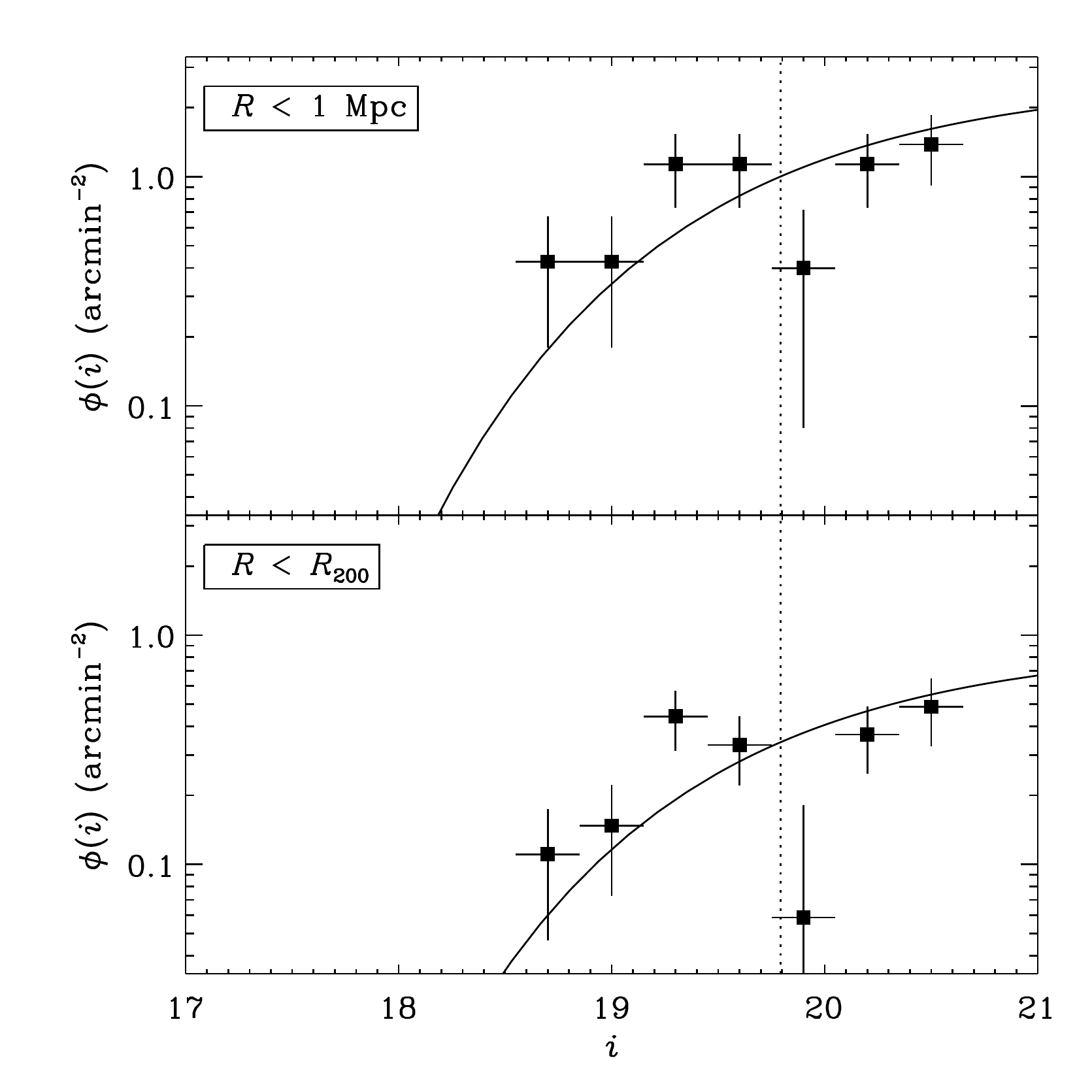}
 \caption{An illustration of luminosity function fitting for SPT-CL
    J0551-5709. We use our passively evolving model to fix $m^*$ to
    $19.77\,\mathrm{mag}$ (vertical dotted line), and integrate down
    to $m^*+1$ to estimate $\ngal$ (top panel) and $\ntwohundred$
    (bottom panel). $\rtwohundred$ is estimated from $\ngal$, as
    described in the text.\label{fig:lf}}
\end{figure}

We stress that the LF fitting step deviates from the MaxBCG procedure,
but it is necessary because our photometry is not complete to $0.4L^*$
on all clusters. 
Our fitting and extrapolating the LF of fairly bright, red-sequence
satellite galaxies is physically reasonable because it is known that
the luminosity function of such members of massive clusters is only
very weakly dependent on mass and is well described by a Schechter
function from lower redshifts \citep{hansen09} to redshifts close to
unity \citep{gilbank08}.
Figure~\ref{fig:lf}
illustrates the resulting LF fit for one of the SPT clusters.

$R_{200}$ is estimated from $\ngal$ using the empirical
$\ngal$--$R_{200}$ relation of \citet{hansen05},
\begin{equation}
\label{eqn:r200ngal}
\rtwohundred = 0.156 \ngal^{0.6}\,h^{-1}\,\mathrm{Mpc}.
\end{equation}
This relation was measured from the MaxBCG cluster sample, which
ranged in redshift from 0.1 to 0.3 and had a median mass of about
$1\times10^{14}\,h^{-1}\,\msun$ .  The clusters we present in this
work are mostly above this redshift range, and expected to
have higher median mass (see V10, \S \ref{sec:richness}).
Nonetheless, we adopt this relation to estimate cluster radii, and
leave verification of the relation on an SZ selected sample such as
this to future work.

The entire richness procedure is repeated, now setting $R=R_{200}$
instead of $1\,h^{-1}\,\mathrm{Mpc}$, to arrive at an estimate of
$\ntwohundred$.  $\ntwohundred$ is then used to estimate cluster mass
using previously established empirical relations, which we now outline.

\subsubsection{Mass-Richness Scaling and Scatter}
\label{sec:scaling}

By comparing to weak gravitational lensing masses,
$\ntwohundred$ richness has been shown by \citet{reyes08} to scale
with $\mtwohundred$, 
the cluster mass contained within a sphere that has an average mass
density of 200 times the universal average,
as
\begin{equation}
  \label{eqn:reyes}
  \frac{\mtwohundred}{10^{14}h^{-1}\,\msun} = (1.42\pm0.08) \left(
    \frac{\ntwohundred}{20} \right)^{1.16\pm 0.09}.
\end{equation}
Earlier, \citet{johnston07} presented a similar, independent weak
lensing study, and their relation (normalization $0.88\pm0.12$, power
law $1.28\pm0.04$) differs from Equation \eqref{eqn:reyes} by about
$30\%$ in mass at $5 \times 10^{14}\,h^{-1}\,\msun$.  We adopt
Equation \eqref{eqn:reyes} as our mass-richness scaling relation with
overall uncertainty of 30\%.

Scatter in $\mtwohundred$ at fixed $\ntwohundred = 40$ as determined
from weak lensing and X-ray cluster masses is $\sigma_{M|N}=45\%\pm
20\%$ \citep[$95\%\,\textrm{CL}$,][]{rozo09}. 

If richness quantities scale as
\begin{equation}
  \mtwohundred \sim \ntwohundred^{1/\alpha} \sim \ngal^{1/(\alpha\beta)} = \ngal^{1/0.56} \sim \rtwohundred^3,
\end{equation}
where $\alpha$ is the $\ntwohundred$--$\mtwohundred$ power law and
$\beta$ is the $\ngal$--$\ntwohundred$ power law, then $\alpha=0.86$
and $\beta=0.65$, using results of Equations~\eqref{eqn:r200ngal} and
\eqref{eqn:reyes}.  Scatter of $45\%\;(60\%)$ in mass translates to
scatter of $39\%\;(52\%)$ in $\ntwohundred$, $25\%\;(34\%)$ in $\ngal$,
and $15\%\;(20\%)$ in $\rtwohundred$.

As with the $\ngal$--$\rtwohundred$ relation (Equation
\eqref{eqn:r200ngal}), these $\ntwohundred$--$\mtwohundred$ mass
relations were determined from weak lensing measurements of MaxBCG
clusters in SDSS, a sample that was deemed to be complete at redshifts
$0.1 < z <0.3$, with median mass of approximately
$1\times10^{14}\,h^{-1}\,\msun$.  Our SZ selected cluster sample has a
significantly broader redshift distribution, as well as a median mass
about five times larger.  Red galaxies are known to be biased tracers
of dark matter, and the bias is a function of mass, redshift, and
radius.  Given the small size of our sample and the large intrinsic
uncertainties inherent to richness techniques, we assume our data are
insensitive to deviations from these fiducial relations, and we leave
the study of the evolution of red galaxy populations with larger SZ
cluster samples to future work.

\subsubsection{Statistical Richness Uncertainties}

We estimated statistical uncertainties on $\ngal$ and $\ntwohundred$
by bootstrapping \citep{efron79} the entire richness procedure
thousands of times. In particular, we assumed that the statistically
limiting step in our procedure was the luminosity function fitting to
the binned magnitude data. Therefore, the start of the bootstrap
process was a random resampling of the magnitude bins themselves. For
each realization, we re-fitted the luminosity function, integrated to
obtain $\ngal$, and estimated $R_{200}$.  We independently
bootstrapped the LF fitting for $\ntwohundred$ estimation as well.
Parameters were assigned as the biweight means \citep{beers90} of
bootstrap distributions, and sample uncertainties as the biweight
standard deviations.

\subsection{Spectroscopic Redshifts}

For each galaxy spectrum, the redshift was found by cross-correlating
with the {\sc fabtemp97} template, using the RVSAO package in IRAF
\citep{kurtz98}. The validity of the cross-correlation redshift was
checked by visual inspection and judged by the presence of visible
absorption (and in a few cases, emission) lines.  Redshift
uncertainties were estimated as two times those given by RVSAO for the
BCG redshifts, or the biweight interval estimator \citep{beers90} for
the other cases.  We discard non-galactic spectra as well as redshifts
in strong disagreement with the ensemble average or our prior
photometric redshift estimate.  The redshift adopted for each cluster
is the median redshift of the galaxies passing all cuts.


\section{Results}
\label{sec:results}

\subsection{Redshifts}
\label{sec:redshifts}

\begin{figure}
 \epsscale{1.2}
\plotone{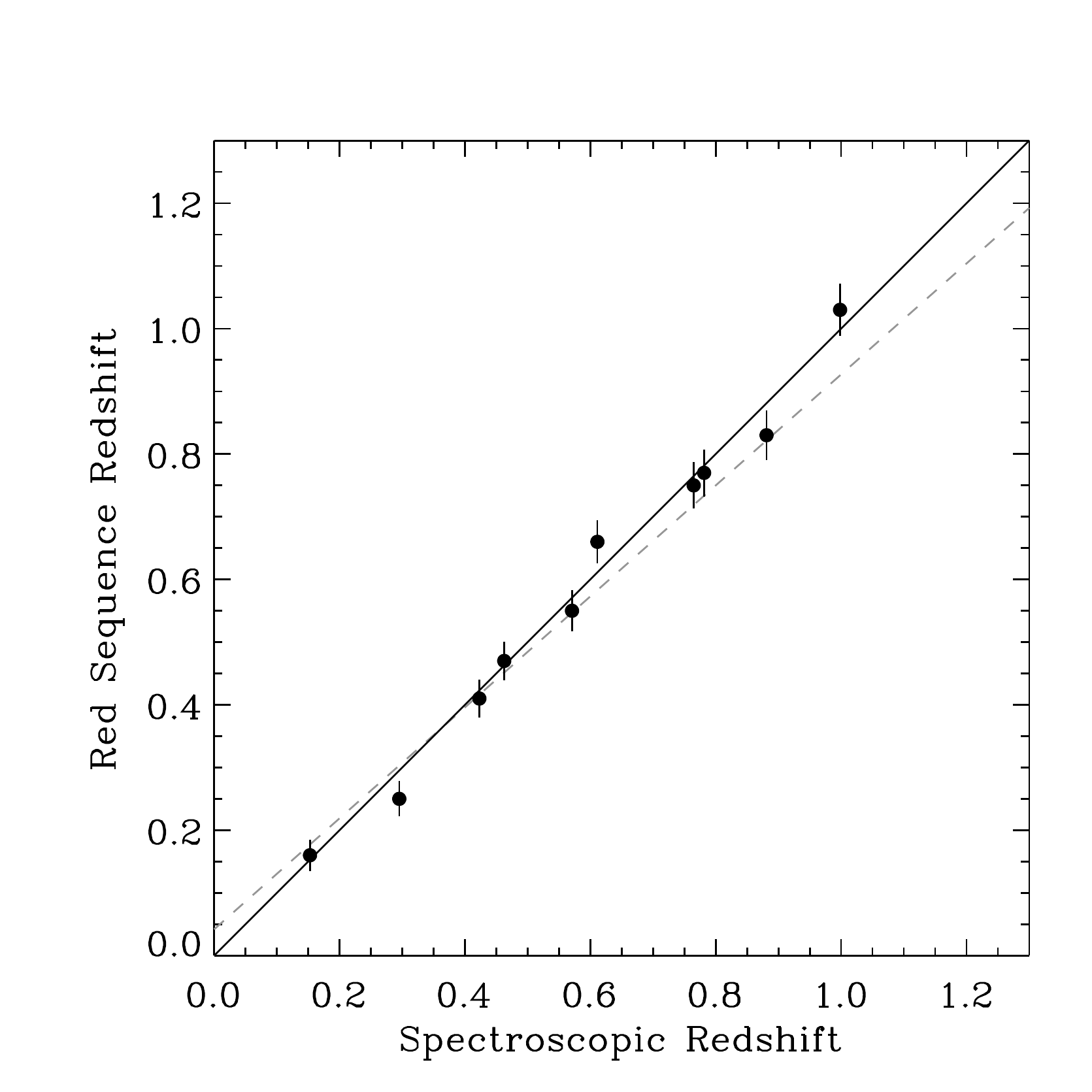}
\caption{Spectroscopic versus red-sequence redshifts. We have applied
  an empirical linear correction to the red-sequence model colors
  using this sample, and this plot shows the result of red-sequence
  redshift measurements after the model correction.  The best fit line
  to the uncorrected redshifts is the dashed line, shown for
  comparison.  Typical RMS redshift scatter is about $2\%$ in
  $\sigma_z/(1+z)$.  Redshift estimates for the entire sample are
  presented in Table \ref{tab:redshift}. \label{fig:zio}}
\end{figure}

We list redshift results in Table~\ref{tab:redshift}.
Figure~\ref{fig:zio} is a plot of red sequence redshifts versus
spectroscopic redshifts for clusters on which we have both
measurements.  The line we fitted to uncorrected redshifts, which were
used to empirically calibrate the red-sequence model colors (\S
\ref{sec:rs}), is the dashed line in the figure.  After the correction, we verify
that the red
sequence redshifts are unbiased to within the
uncertainties.  Root-mean-square scatter in redshift per cluster is
$\sigma_z=0.02(1+z)$, with 
maximum absolute deviation of $0.05(1+z)$.  We therefore
generically assign random uncertainties of $2\%$ to all redshifts in
the range we have directly tested, $0.15<z<1$.  For redshifts
estimated to be $>1$, we assign $5\%$ uncertainties in order to
effectively rule out redshifts less than 1.  Finally, for clusters at
$z<1$ where our photometric uncertainties are large, we also
generically increase the uncertainty estimates.


\ifthenelse{\equal{\ispreprint}{true}}
{
\begin{deluxetable*}{lllccc}
}
{
\begin{deluxetable}{lllccc}
}
\tabletypesize{\scriptsize}
\tablecaption{Cluster Redshift Data\label{tab:redshift}}
\tablewidth{0pt}
\tablehead{
  \colhead{Cluster Name} &
  \colhead{$\redseqz$} & 
  \colhead{$\specz$} &
  \colhead{$N_{\mathrm{spec}}$} &
  \multicolumn{2}{c}{Imaging Coverage} \\
  ~ & ~ & ~ & ~ &
  \colhead{BCS?} &
  \colhead{Magellan?}
}
\startdata
SPT-CL J0509-5342 &  0.47(3) &   0.4626(4) &        6 & Y & Y \\ 
SPT-CL J0511-5154 &  0.74(3) &     \nodata &  \nodata & N & Y \\ 
SPT-CL J0516-5430 &  0.25(3) &      0.2952 &        8 & Y & Y \\ 
SPT-CL J0521-5104 &  0.72(3) &     \nodata &  \nodata & Y & N \\ 
SPT-CL J0528-5259 &  0.75(4) &   0.7648(5) &        2 & Y & Y \\
SPT-CL J0533-5005 &  0.83(4) &   0.8810(9) &        4 & N & Y \\
SPT-CL J0539-5744 &  0.77(4) &     \nodata &  \nodata & N & Y \\
SPT-CL J0546-5345 & 1.16(11) &  \nodata &       \nodata & Y & N \\
SPT-CL J0551-5709 &  0.41(3) &  0.4230(10) &        5 & N & Y \\
SPT-CL J0559-5249 &  0.66(3) &   0.6112(3) &        5 & N & Y \\ 
SPT-CL J2259-5617 &  0.16(2) &      0.1528 &        1 & N & Y \\ 
SPT-CL J2300-5331 &  0.29(3) &     \nodata &  \nodata & N & Y \\
SPT-CL J2301-5546 &  0.78(9) &     \nodata &  \nodata & N & Y \\
SPT-CL J2331-5051 &  0.55(3) &   0.5707(5) &        8  & N & Y \\
SPT-CL J2332-5358 &  0.32(3) &     \nodata &  \nodata & Y & Y \\
SPT-CL J2337-5942 &  0.77(4) &   0.7814(5) &        2  & N & Y \\ 
SPT-CL J2341-5119 &  1.03(4) &   0.9983(5) &        1  & N & Y \\ 
SPT-CL J2342-5411 & 1.08(10) &     \nodata &  \nodata  & Y & N \\  
SPT-CL J2332-5521 & \nodata &     \nodata &  \nodata  & Y & Y \\  
SPT-CL J2355-5056 &  0.35(3) &     \nodata &  \nodata  & N & Y \\ 
SPT-CL J2359-5009 &  0.76(4) &     \nodata &  \nodata  & N & Y \\ 
SPT-CL J0000-5748 &  0.74(9) &     \nodata &  \nodata  & N & Y
\enddata
\tablecomments{See \S \ref{sec:clusters} for notes on individual
  clusters, including those which have been identified in other
  works.}
\ifthenelse{\equal{\ispreprint}{true}}
{
\end{deluxetable*}
}
{
\end{deluxetable}
}

\subsection{Richness}
\label{sec:richness}

In Figure \ref{fig:n200} we compare our richness-derived masses to
those presented in V10, which were estimated from SPT millimeter-wavelength data.  We
label V10 masses as $M(\xi)$, because they are calculated from the SZ
signal-to-noise ratio $\xi$.  The purpose of 
the comparison is to assess the level of correlation and explore
whether the power law and normalization of the scaling laws are
consistent with previous work.  We additionally estimate the
normalization of $\ngal$, which we treat as an empirical mass
observable in its own right rather than only as a measurement
intermediate to $\ntwohundred$.

\ifthenelse{\equal{\ispreprint}{true}}
{
\begin{figure*}
}
{
\begin{figure}
}
 \includegraphics[scale=0.8]{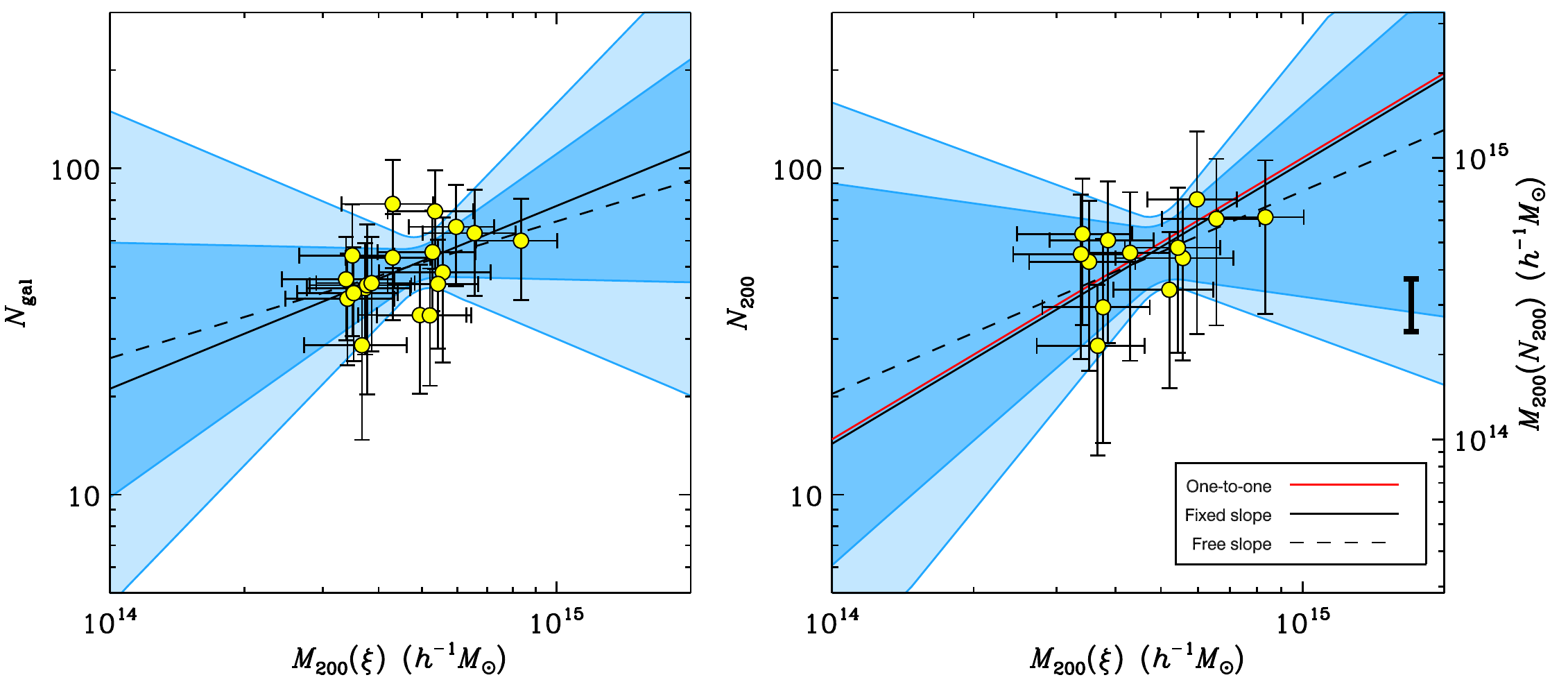}
\caption{Optical richness versus total cluster mass estimated from
   the SPT data. This plot shows that the richness correlates
   highly with this millimeter-wavelength mass-observable \citep[taken
   from][]{vanderlinde10}, and together the data agree with previously
   established scaling relations to within the uncertainties.  The
   solid red line in the right-hand panel is the one-to-one mass
   relation, and the solid black lines in both panels are best-fit
   relations when fixing the slope to previously measured values.  The
   dashed lines are best-fit relations leaving both slope and
   intercept free.  The dark, inner shaded areas denote the $68\%$
   confidence regions assuming zero intrinsic richness-mass scatter
   for the two parameter regression. The light, outer shaded
   areas denote $68\%$ confidence regions assuming $35\%$ scatter in
   mass, which is a better match to the data.  The heavy error bar on
   the far right-hand side indicates the nominal overall uncertainty of
   $30\%$ in the $\ntwohundred$--$\mtwohundred$
   relation.\label{fig:n200}}
\ifthenelse{\equal{\ispreprint}{true}}
{
\end{figure*}
}
{
\end{figure}
}

Richness and $M(\xi)$ are plotted against one another in
Figure~\ref{fig:n200}.  We have used Equation~\eqref{eqn:reyes} to
render the far right-hand mass axis, with the $30\%$ overall
uncertainty in this relation denoted with the heavy error bar.  We do
not display a mass axis in the $\ngal$ panel of the figure because
there have been no previous measurements of an $\ngal$--$\mtwohundred$
scaling relation.

We perform the analysis 
only considering clusters whose radius ($\rtwohundred$ in the case of
$\ntwohundred$, $1\,h^{-1}\,\mathrm{Mpc}$ in the case of $\ngal$) falls fully
within the observed field of view.  For Magellan IMACS imaging this
limited radii to $\lesssim 6\arcmin$, because we placed the SPT center
in the middle of one of the eight chips, at a distance of about six
arcminutes from the field edge.  Spatial incompleteness of this kind
does not affect the BCS data.  

\subsubsection{$\ngal$ Scaling}

$\rtwohundred$ for massive clusters is approximately
$1.5\,h^{-1}\,\mathrm{Mpc}$, which subtends $\gtrsim6\arcmin$ at redshifts
$z\lesssim0.3$.  We find it useful to count red galaxy overdensities
in a smaller region, corresponding to $1\,h^{-1}\,\mathrm{Mpc}$ in co-moving
coordinates, which subtends $\gtrsim6\arcmin$ at redshifts
$z\lesssim0.17$.  This is a better match to the observed typical
angular size of the SZ signal.  In addition, the
surface density of red galaxies in the smaller aperture is greater
than for $\rtwohundred$, whose larger area effectively dilutes the
signal.  For these reasons we explore $\ngal$ as an empirical mass
proxy.

As argued in \S~\ref{sec:scaling}, $\ngal$ should scale with mass as
$\mtwohundred^{0.56}$.  We fix this slope and measure
\begin{equation}
  \label{eqn:ngalm200}
 \ngal = 
  (52 \pm 3 \pm 9) \left( \frac{\mtwohundred}{5\times10^{14}\,h^{-1}\,\msun} \right)^{0.56},
\end{equation}
If zero intrinsic scatter is assumed, the reduced chi-square of the
fit is $\chi^2_\nu\approx3$, whereas $20\%$ in $\ngal$ (corresponding
to $\sim35\%$ in mass)  produces $\chi^2_\nu = 1$. The
first error given in Equation \eqref{eqn:ngalm200} is random-only, which
includes the intrinsic mass-richness scatter, and the second error is
the overall systematic uncertainty of $30\%$.  This fit is shown as
the solid black line in the left-hand panel of Figure~\ref{fig:n200}.

Letting the slope be free, we measure a normalization consistent with
Equation \eqref{eqn:ngalm200} and slope $0.42 \pm 0.50$.
This is the dashed line
in the left-hand panel of Figure~\ref{fig:n200}.  In the figure, the
inner, dark shaded area denotes the $68\%$ confidence region for the
best two-parameter fit assuming zero intrinsic scatter, while the
outer, light shaded area is the $68\%$ confidence region for best
two-parameter fit assuming $35\%$ intrinsic mass scatter, which is a
significantly better match to the data.  We estimate the uncertainty
on intrinsic scatter by varying it until the chi-square doubles,
resulting in relative $1\sigma$ uncertainty at the $\sim50\%$ level.

\subsubsection{$\ntwohundred$ Scaling}
 
We perform the same analysis on $\ntwohundred$.  Fixing the mass-mass
slope to unity, we measure
\begin{equation}
  \label{eqn:n200m200}
  \ntwohundred = 
  (57 \pm 4 \pm 15) \left( \frac{\mtwohundred}{5\times10^{14}\,h^{-1}\,\msun} \right)^{0.86}.
\end{equation}
Again, $35\%$ intrinsic mass scatter, corresponding to $30\%$ scatter
in $\ntwohundred$,
gives $\chi^2_\nu=1$.  This fit is
shown as the black line in right-hand panel of Figure \ref{fig:n200},
and can be compared to the one-to-one mass line, in red.

Letting the slope be free, the measured normalization consistent with
Equation \eqref{eqn:n200m200} and slope $0.62 \pm 0.73$.
This is shown
with the dashed black line in Figure~\ref{fig:n200}.  As before, the
dark, inner shaded region is the $68\%$ confidence region for the best
two-parameter fit assuming zero intrinsic scatter, while the outer,
light shaded area is the $68\%$ confidence region for best
two-parameter fit assuming $35\%$ intrinsic mass scatter, which is a
significantly better match to the data.  Relative uncertainty on the
intrinsic scatter is comparable to that quoted above for $\ngal$.

All richness results are given in Table \ref{tab:richness}.  In the
table we have adopted Equation \eqref{eqn:reyes} to estimate masses
from $\ntwohundred$, and for $\ngal$ masses we have used Equation
\eqref{eqn:ngalm200}.  Systematic uncertainties are taken to be the
quadrature sum of the nominal $45\%$ intrinsic scatter in mass (\S
\ref{sec:scaling}) and the $30\%$ overall uncertainty in the
richness--mass scaling relation.  Further work on a larger sample of
clusters selected with similar criteria as these is needed to reduce
statistical uncertainties and measure the scatter directly.  We
discuss the implications of our richness measurements in \S
\ref{sec:discussion}.


\ifthenelse{\equal{\ispreprint}{true}}
{
\begin{deluxetable*}{llrllr}
}
{
\begin{deluxetable}{llrllr}
}
\tabletypesize{\scriptsize}
\tablecaption{Cluster Richness Data\label{tab:richness}}
\tablewidth{0pt}
\tablehead{
  \colhead{Object Name} &
 \colhead{$\ngal$\tablenotemark{a}} &
  \colhead{$\mtwohundred(\ngal)$\tablenotemark{b}} &
  \colhead{$\rtwohundred$\tablenotemark{a}} &
 \colhead{$\ntwohundred$\tablenotemark{a}} & 
  \colhead{$\mtwohundred(\ntwohundred)$\tablenotemark{b}} \\
~ & ~  & \colhead{($10^{14}h^{-1}\,\msun$)} & \colhead{($\,h^{-1}\mathrm{Mpc}$)}  & ~ & \colhead{($10^{14}h^{-1}\,\msun$)}
}
\startdata
SPT-CL J0509-5342 & $ 41( 8)$ & $ 3.3\pm2.0\pm1.0$ & $1.46(17)$ & $ 51(13)$ & $ 4.3\pm2.5\pm1.3$ \\ 
SPT-CL J0511-5154 & $ 77(12)$ & $10.3\pm5.9\pm3.1$ & $2.13(20)$ &  \nodata &         \nodata \\ 
SPT-CL J0516-5430 &  \nodata &         \nodata &  \nodata &  \nodata &         \nodata \\ 
SPT-CL J0521-5104 & $ 44( 7)$ & $ 3.7\pm2.2\pm1.1$ & $1.52(15)$ & $ 57(13)$ & $ 4.8\pm2.7\pm1.4$ \\ 
SPT-CL J0528-5300 & $ 44( 9)$ & $ 3.8\pm2.3\pm1.1$ & $1.52(18)$ & $ 60(13)$ & $ 5.1\pm2.8\pm1.5$ \\ 
SPT-CL J0533-5005 & $ 28(10)$ & $ 1.7\pm1.4\pm0.5$ & $1.17(25)$ & $ 28( 6)$ & $ 2.2\pm1.2\pm0.6$ \\ 
SPT-CL J0539-5744 & $ 63( 9)$ & $ 7.1\pm4.0\pm2.1$ & $1.88(16)$ & $ 69(14)$ & $ 6.1\pm3.3\pm1.8$ \\ 
SPT-CL J0546-5345 & $ 66( 7)$ & $ 7.7\pm4.1\pm2.3$ & $1.93(12)$ & $ 80(31)$ & $ 7.1\pm4.8\pm2.1$ \\ 
SPT-CL J0551-5709 & $ 54(15)$ & $ 5.4\pm3.8\pm1.6$ & $1.71(29)$ &  \nodata &         \nodata \\ 
SPT-CL J0559-5249 & $ 59( 6)$ & $ 6.5\pm3.4\pm1.9$ & $1.82(11)$ & $ 70( 8)$ & $ 6.2\pm3.2\pm1.8$ \\ 
SPT-CL J2259-5617 &  \nodata &         \nodata &  \nodata &  \nodata &         \nodata \\ 
SPT-CL J2300-5331 & $ 35(10)$ & $ 2.5\pm1.8\pm0.8$ & $1.33(22)$ &  \nodata &         \nodata \\ 
SPT-CL J2301-5546 & $ 35( 8)$ & $ 2.5\pm1.6\pm0.8$ & $1.33(17)$ & $ 42( 6)$ & $ 3.4\pm1.8\pm1.0$ \\ 
SPT-CL J2331-5051 & $ 73( 5)$ & $ 9.4\pm4.8\pm2.8$ & $2.06( 8)$ &  \nodata &         \nodata \\ 
SPT-CL J2332-5358 & $ 42( 8)$ & $ 3.5\pm2.1\pm1.1$ & $1.49(16)$ &  \nodata &         \nodata \\ 
SPT-CL J2337-5942 & $ 53( 8)$ & $ 5.2\pm2.9\pm1.6$ & $1.69(15)$ & $ 55(11)$ & $ 4.6\pm2.5\pm1.4$ \\ 
SPT-CL J2341-5119 & $ 39( 7)$ & $ 3.1\pm1.9\pm0.9$ & $1.42(16)$ & $ 62( 7)$ & $ 5.4\pm2.8\pm1.6$ \\ 
SPT-CL J2342-5411 & $ 43(19)$ & $ 3.7\pm3.4\pm1.1$ & $1.51(38)$ & $ 37(12)$ & $ 2.9\pm1.8\pm0.9$ \\ 
SPT-CL J2355-5056 & $ 55( 5)$ & $ 5.6\pm3.0\pm1.7$ & $1.73(10)$ &  \nodata &         \nodata \\ 
SPT-CL J2359-5009 & $ 47(16)$ & $ 4.3\pm3.4\pm1.3$ & $1.59(32)$ & $ 53( 8)$ & $ 4.4\pm2.3\pm1.3$ \\  
SPT-CL J0000-5748 & $ 45( 6)$ & $ 4.0\pm2.2\pm1.2$ & $1.55(12)$ & $ 54(11)$ & $ 4.6\pm2.5\pm1.4$
\enddata
\tablecomments{}
\tablenotetext{a}{Uncertainties given are statistical only.}
\tablenotetext{b}{Uncertainties given are statistical and systematic, respectively.}
\ifthenelse{\equal{\ispreprint}{true}}
{
\end{deluxetable*}
}
{
\end{deluxetable}
}

\subsection{Notable Clusters}
\label{sec:clusters}

In this section we describe notable information, if any, about the
clusters.  As we will point out, a subset of clusters also appear in
the catalogs of \citetalias{abell89}, \citet{boehringer04},
\citetalias{staniszewski08}, \citet[][hereafter
\citetalias{menanteau09}]{menanteau09}, \citet[][hereafter
\citetalias{menanteau08}]{menanteau08},  and \citet[][hereafter
\citetalias{menanteau10}]{menanteau10}.  Our redshifts agree to within
$\sigma_z \approx 3\%(1+z)$ of the photometric redshifts presented in
\citetalias{menanteau08} and \citetalias{menanteau10}, except for the
highest redshift cluster from \citetalias{staniszewski08}, SPT-CL
J0546-5345 (see below).  Because we have not presented exhaustive
optical cluster-finding in the entire BCS survey in this work, instead
having concentrated on fields in the direction of the SZ detections,
and because the number of overlapping clusters is too small to draw
useful conclusions with high statistical confidence, we leave a formal
inter-comparison of redshift and richness results of
\citetalias{menanteau08} and \citetalias{menanteau10} to future
studies.

\paragraph{SPT-CL J0509-5342}

This cluster was previously identified by \citetalias{staniszewski08}.  

\paragraph{SPT-CL J0511-5154} 

This cluster has recently been identified by \citetalias{menanteau10},
who assigned the name SCSO J051145-515430.

\paragraph{SPT-CL J0516-5430}

This cluster was previously identified by \citetalias{staniszewski08},
where it was called by a different name, SPT-CL J0517-5430.  The SPT
name ascribed to this object in this work and in V10 follow the
recommendations of the International Astronomical Union (IAU), and
should be adopted permanently.  This cluster also identified as Abell
S0520 \citep[][hereafter A89]{abell89}, and RXCJ0516.6-5430
\citep{boehringer04}, the latter of which is the source of the
spectroscopic redshift.  \citetalias{menanteau10} detected this object
and call it SCSO J051637-543001.

\paragraph{SPT-CL J0521-5104}  

This cluster has been identified by \citetalias{menanteau10} as SCSO
J052113-510418.

\paragraph{SPT-CL J0528-5300}

This cluster was previously identified by \citetalias{staniszewski08},
where it was called by the same name, and was also identified by
\citetalias{menanteau10} as SCSO J052803-525945.


\paragraph{SPT-CL J0539-5744} This cluster displays a possible strong
gravitational lens arc.

\paragraph{SPT-CL J0546-5345}

This cluster was previously identified by \citetalias{staniszewski08},
where it was called by a different name, SPT-CL J0547-5345.  The SPT name
ascribed to this object in this work and in V10 follow the
recommendations of the IAU, and should be adopted permanently.
\citetalias{staniszewski08} reported a photometric redshift of
$\sim0.9$ and \citetalias{menanteau08} independently reported 
$z_{\mathrm{photo}}=0.88^{+0.08}_{-0.04}$.  We do not detect
a red-sequence overdensity near redshift $0.9$, but a faint
red-sequence peak is
evident at $\redseqz \approx 1.15$. 

\paragraph{SPT-CL J0551-5709}

Abell S0552 \citepalias{abell89} is in the foreground of this cluster.  No redshift
estimate exists for Abell S0552, but a strong red sequence at $\redseqz
= 0.09$ is clearly visible in color-magnitude diagrams.  The SPT
cluster identified here, however, is measured at $z = 0.42$.

\paragraph{SPT-CL J0559-5249} 

Our red-sequence studies reveal two
significant red galaxy overdensities, one at the redshift given in
Table \ref{tab:redshift} and another at $z\approx0.4$.  On inspection
of the spatial distribution of galaxies, we attribute the SZ signal
to the higher redshift system.

\paragraph{SPT-CL J2259-5617}

We identify this cluster with Abell 3950, and recover a spectroscopic
redshift from archival data on what we identify as the BCG.  The SPT
SZ detection coordinate lies nearly exactly on the line joining the
two Abell 3950 coordinates given in the literature, at a projected
distance of $71\arcsec$ from that quoted by \citet[][]{arp96}, and
$208\arcsec$ from that quoted by A89.  The redshift of Abell 3950 has
not been previously measured, but we have identified the BCG in 2MASS
as 2MASX J23000108-5617061, which the 6dF Galaxy Survey measured to be
at $\specz=0.152787$ \citep{jones05b}.  This galaxy lies $18\arcsec$
from the SPT coordinate.

\paragraph{SPT-CL J2300-5331} 

We identify this with Abell S1079 (A89).  No previous redshift
estimates exist for this cluster.


\paragraph{SPT-CL J2331-5051} 

This cluster exhibits a giant gravitational lens arc and a well
separated secondary cluster structure in both the optical and SZ
data.  This is among the most interesting of the clusters presented
here, and is the subject of a dedicated study (High et al., in
preparation).

\paragraph{SPT-CL J2332-5358} 

This cluster was recently identified by \citetalias{menanteau10} as
SCSO J233227-535827.




\paragraph{SPT-CL J2343-5521} 

No red-sequence cluster appears in BCS imaging, whose 50\%
completeness depth in the $i$ band is 23.5 mag, corresponding to
$m^{*}$ at $z \approx 1.2$.  Either there exists a cluster at higher
redshift, or this is a false SZ detection.  V10 show that the false
detection rate is only 7\% for clusters at equivalent SZ significance.
However, the SZ profile radius is significantly larger than any other
cluster, consistent with a cosmic microwave background fluctuation.
Preliminary follow-up of this cluster at other wavelengths suggest
this is indeed a false detection.





\section{Discussion of Systematic Effects}
\label{sec:discussion}

The greatly different criteria with which our SZ clusters were
selected as compared to the MaxBCG sample could give rise to
differences in measured properties.  
One important effect is the evolution of the mean and scatter of red
cluster-galaxy colors with redshift and mass.  While elliptical (E)
galaxies are highly homogeneous in the range of redshifts our sample
represents, $0 < z \lesssim 1.2$ \citep{menci08,lidman08,mei09}, the
mean and scatter of S0 galaxy colors are known to evolve with redshift
and density of the environment \citep[e.g.,][]{vandokkum98}.  The
environment also causes color evolution with distance from the cluster
center.  Our method, and indeed that of MaxBCG, does not explicitly
use morphological selection criteria, so we must assume E and S0
populations contribute to our richness estimates.  Understanding 
color-selection effects at all redshifts for a large sample of
clusters would require sophisticated simulations or full photometric
redshift measurements that are beyond the scope of this work.  

Another important effect is the evolution of the abundance of
early-type cluster galaxies as a function of luminosity.  Clusters
have been observed to accumulate faint galaxies at a greater rate than
bright ones over time, manifesting as evolution in the slope of the
luminosity function's faint end, $\alpha$, as a function of redshift
\citep{rudnik09}.  Despite our long redshift baseline, we expect our
measurements here to be largely unaffected by behavior at the faint
end, as we integrate the LF down to a relatively shallow magnitude of
$m^*+1$.  Indeed, tests where we vary $\alpha$ have not significantly
affected our results.

In addition, because our richness measurement is not exactly that used
with maxBCG (from which the $\rtwohundred$--$\ngal$ relationship that
we use here is taken) our $\rtwohundred$ estimates are likely to be
somewhat overestimated \citep[cf.\, discussion in][]{hansen09}. Using a
larger-than-ideal aperture for counting red galaxies may contribute to
some of the scatter in the richness-mass correlation that we observe
here.



Red galaxy counting has nonetheless proven to be a simple and
accessible way to estimate the total mass in clusters and groups.  If
this technique can be accurately extended to the very wide range of
redshifts that SPT SZ-selected clusters span, then modeling and
measuring evolutionary effects in 
SZ clusters will be useful to obtaining constraints on
masses of very large cluster samples.


\section{Conclusions}\label{sec:conclusion}

We have observed clusters from the 2008 South Pole Telescope SZ survey
at optical and near-infrared wavelengths.  We estimate redshifts and
richness with red-sequence techniques, and we obtain spectroscopic
redshifts for a subsample of the clusters.

Our red-sequence-derived redshifts exhibit $2\%$ RMS scatter in
$\sigma_z/(1+z)$ in the subsample with spectroscopic overlap, over the
redshift range $0.15<z<1.0$.  
Our analysis provides no evidence that the SZ selected sample from SPT
follows different scaling relations than those followed by SDSS
optically selected clusters.

The clusters presented in this paper comprise the largest sample of
galaxy clusters discovered with the SZ and demonstrates that current
SZ surveys can detect many high-mass galaxy clusters across a wide
range of redshifts.  Precise dark energy 
constraints from these surveys require the cluster redshifts,
masses, and selection function to be known.  The SZ effect contains no redshift information,
and coordinated observations at optical and infrared wavelengths are
an efficient means of providing this, especially for large cluster
samples.  Optical cluster identification is also potentially useful for
understanding SZ selection functions.
In the future, there will probably be greater coverage of SZ
clusters by OIR imaging than any other wavelength or observing method
other than the millimeter-wavelength itself.  Optical redshifts are essential for
constraining cosmology with SZ surveys, and the same data may also be
brought to bear on the mass and cluster selection problems. 

\acknowledgments

This research has made use of the NASA/IPAC Extragalactic Database
(NED) which is operated by the Jet Propulsion Laboratory, California
Institute of Technology, under contract with the National Aeronautics
and Space Administration.  This publication has made use of data
products from the Two Micron All Sky Survey, which is a joint project
of the University of Massachusetts and the Infrared Processing and
Analysis Center/California Institute of Technology, funded by the
National Aeronautics and Space Administration and the National Science
Foundation.  This research has made use of the NASA/ IPAC Infrared
Science Archive, which is operated by the Jet Propulsion Laboratory,
California Institute of Technology, under contract with the National
Aeronautics and Space Administration.

IRAF is the Image Reduction and Analysis Facility, distributed by the
National Optical Astronomy Observatories, which are operated by the
Association of Universities for Research in Astronomy, Inc., under
cooperative agreement with the National Science Foundation.

We used the cosmology calculator tools of \citet{wright06}.

This work is supported by the NSF (AST-0607485, AST-0506752,
ANT-0638937, ANT-0130612, MRI-0723073), the DOE (DE-FG02-08ER41569 and
DE-AC02-05CH11231), NIST (70NANB8H8007), and Harvard University.
B.\ Stalder and A.\ Loehr gratefully acknowledge support by the
Brinson Foundation.  R.\ J.\ Foley acknowledges the generous support
of a Clay fellowship.

We also thank the team of
scientists, engineers and observing staff of the Las Campanas
Observatory and of Cerro Tololo Inter-American Observatory.

{\it Facilities:}
\facility{Magellan:Baade (\imacs)},
\facility{Magellan:Clay (\ldss)},
\facility{CTIO:Blanco (MosaicII)},
\facility{South Pole Telescope}

\bibliographystyle{fapj}
\bibliography{ms}



\ifthenelse{\equal{\ispreprint}{true}}
{

  \appendix

  \input{thumbs_appendix}

}
{

}

\end{document}

%% file: thumbs_appendix.tex
\section{Gallery of Clusters}
\label{app:thumbs}

\begin{figure}
  \plotone{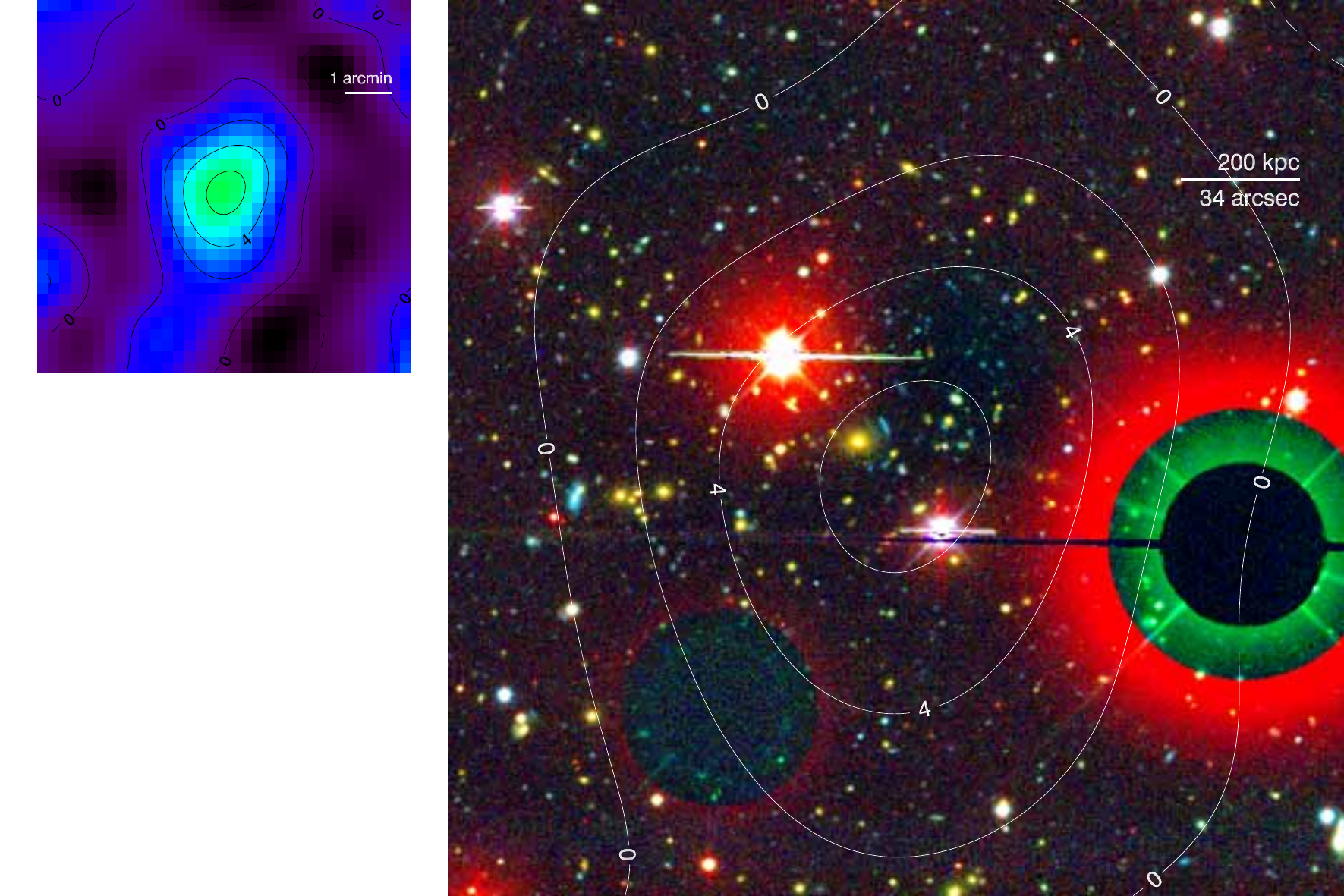}
 \caption{SPT-CL J0509-5342\label{fig:thumb1}}
\end{figure}

\begin{figure}
  \plotone{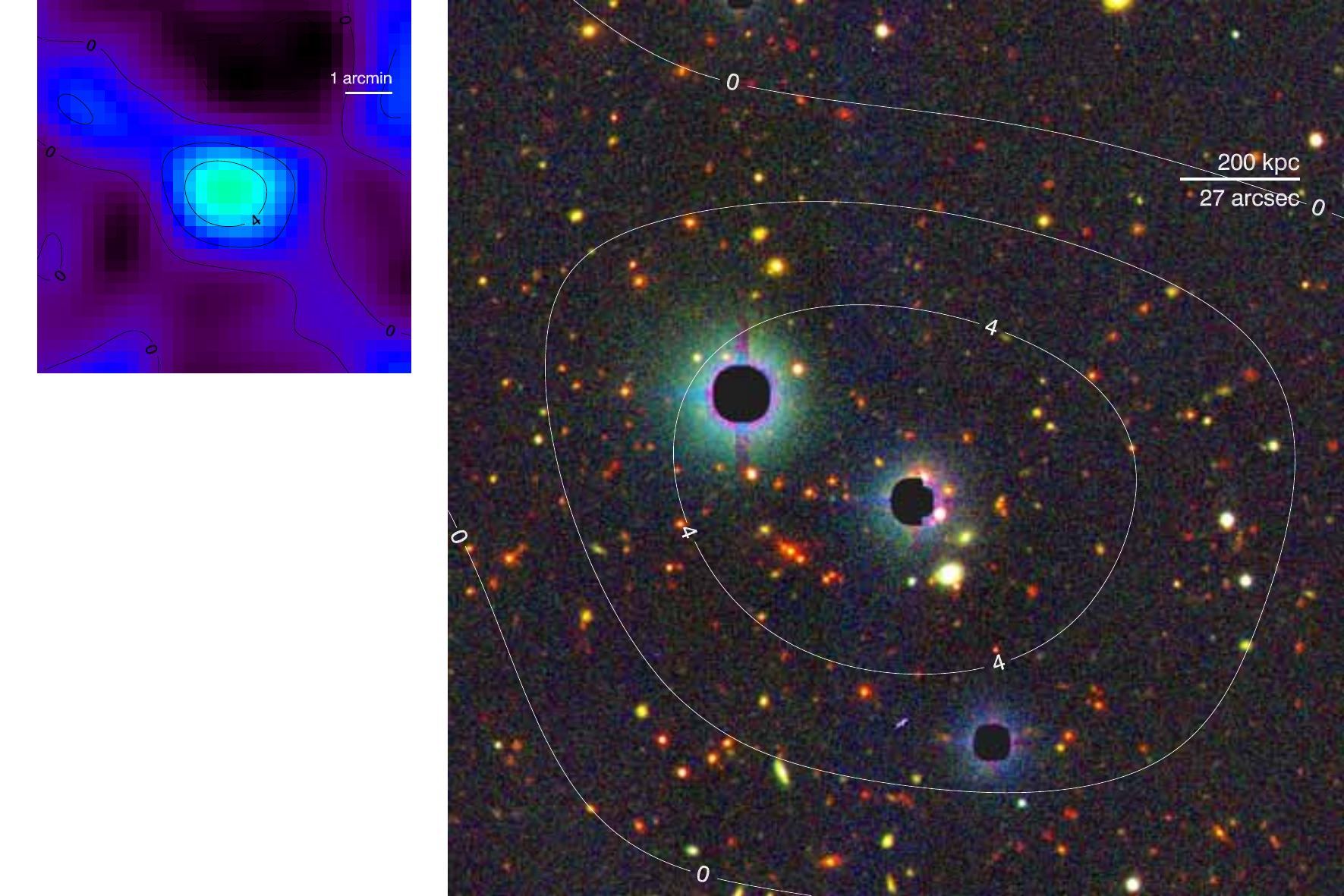}
 \caption{SPT-CL J0511-5154\label{fig:thumb2}}
\end{figure}

\begin{figure}
  \plotone{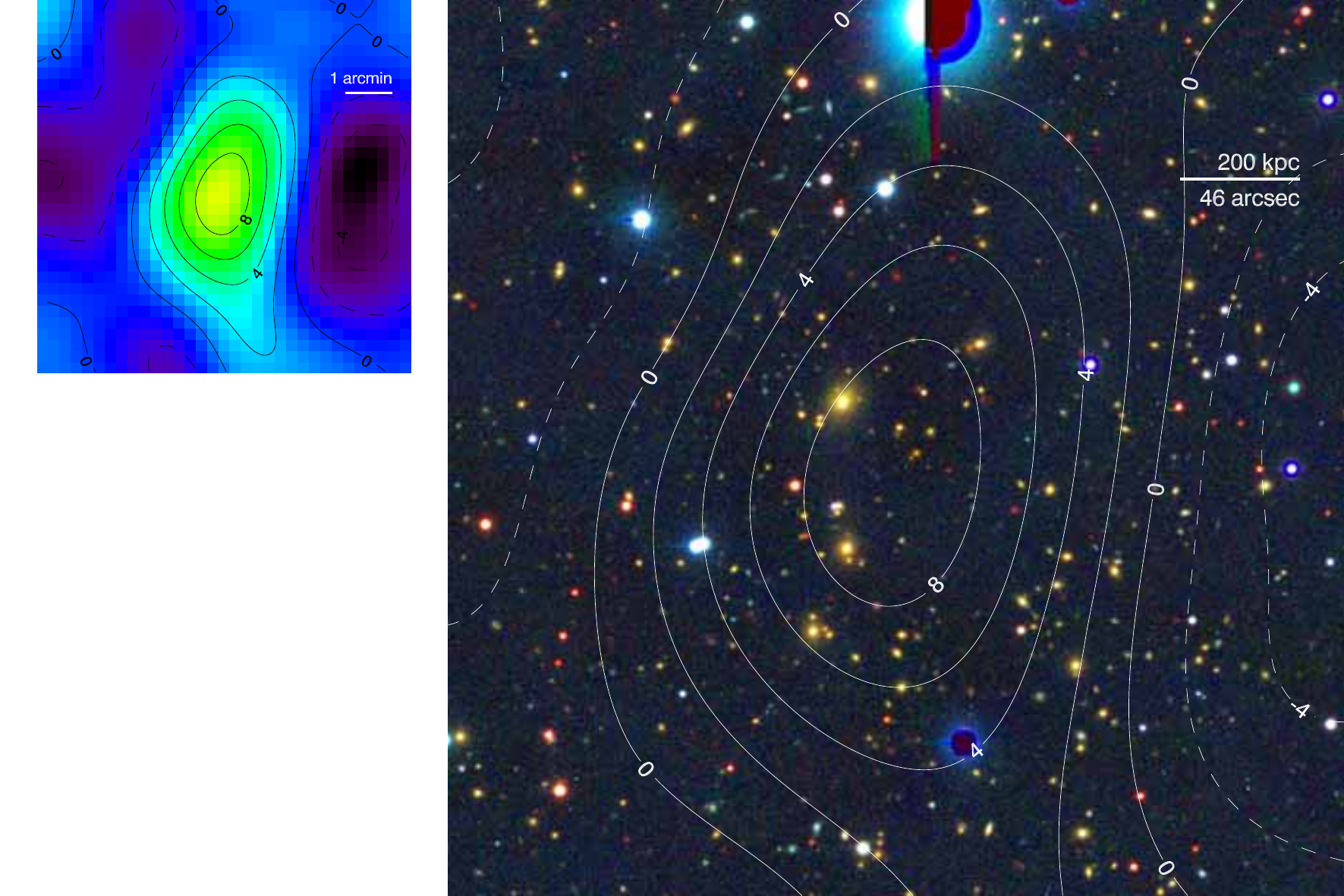}
 \caption{SPT-CL J0516-5430\label{fig:thumb3}}
\end{figure}

\begin{figure}
  \plotone{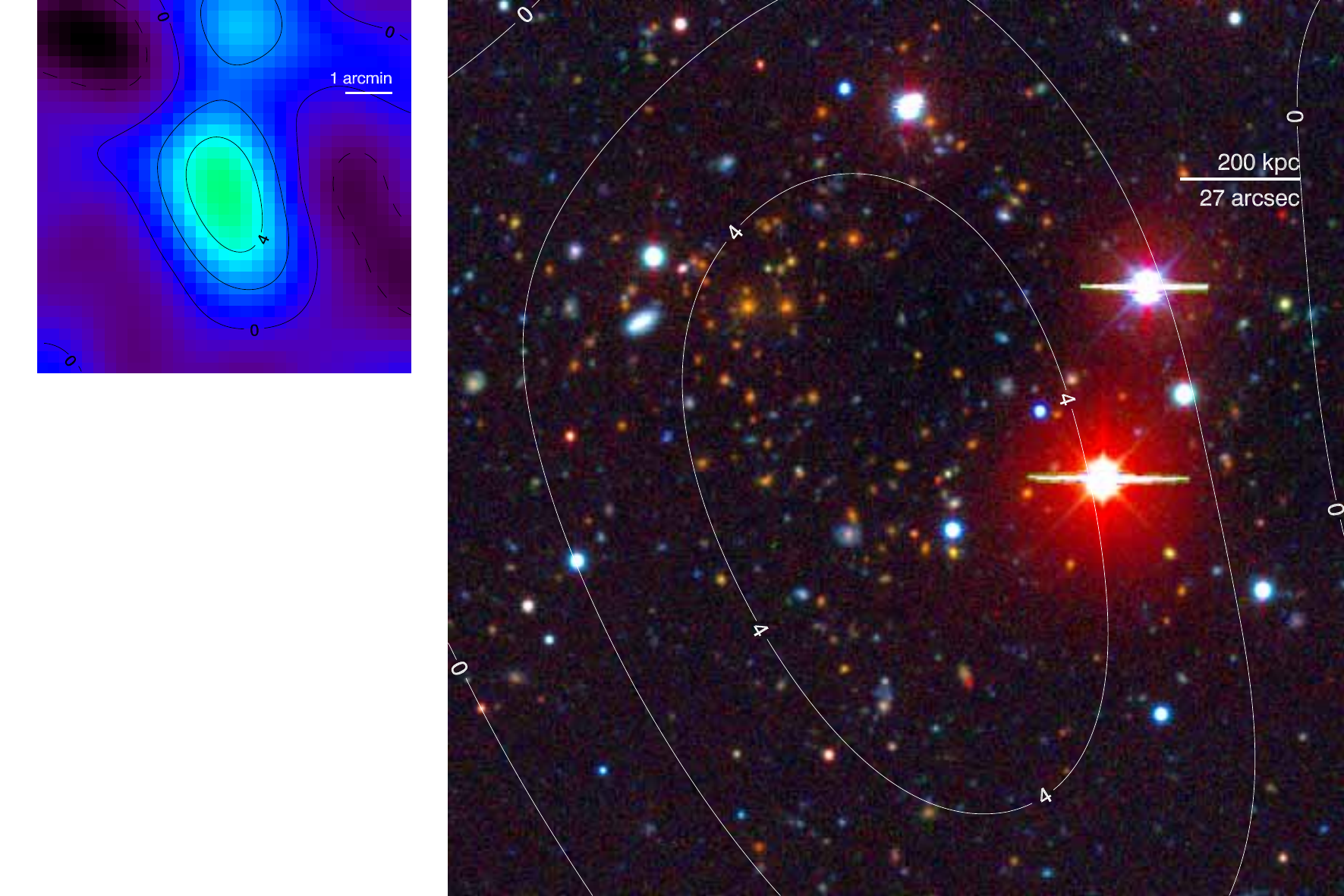}
 \caption{SPT-CL J0521-5104\label{fig:thumb4}}
\end{figure}

\begin{figure}
  \plotone{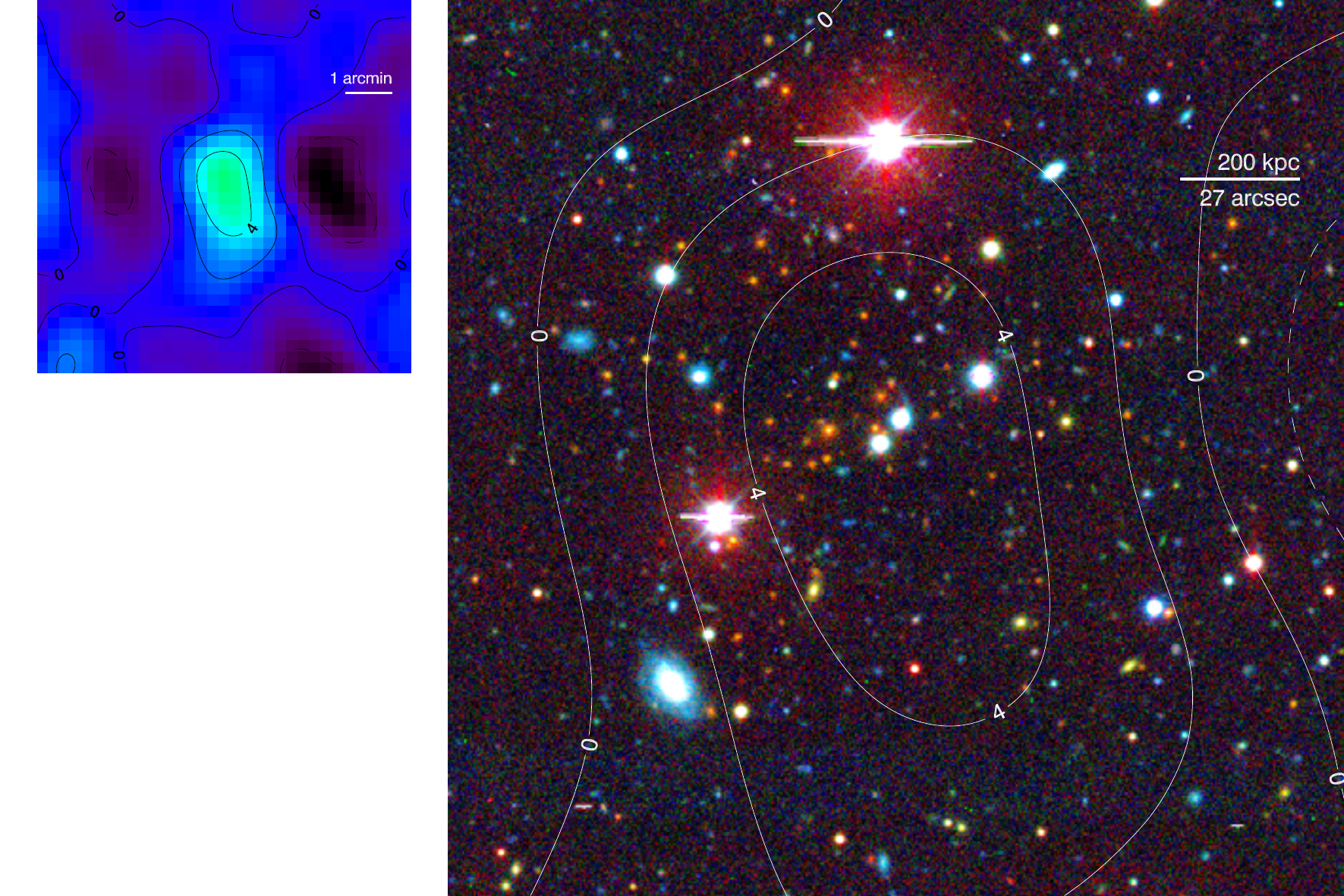}
 \caption{SPT-CL J0528-5300\label{fig:thumb5}}
\end{figure}

\begin{figure}
  \plotone{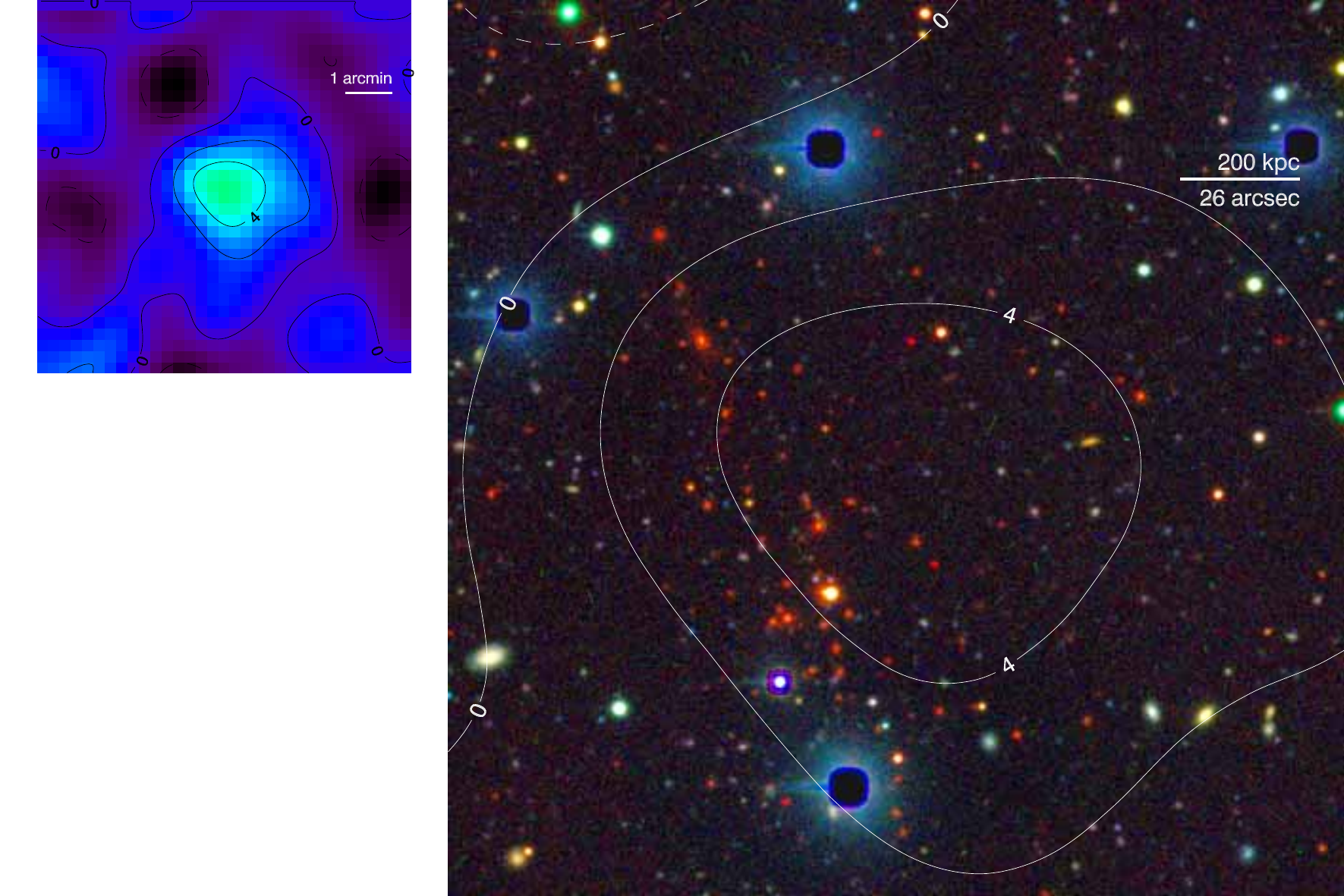}
 \caption{SPT-CL J0533-5005\label{fig:thumb6}}
\end{figure}

\begin{figure}
  \plotone{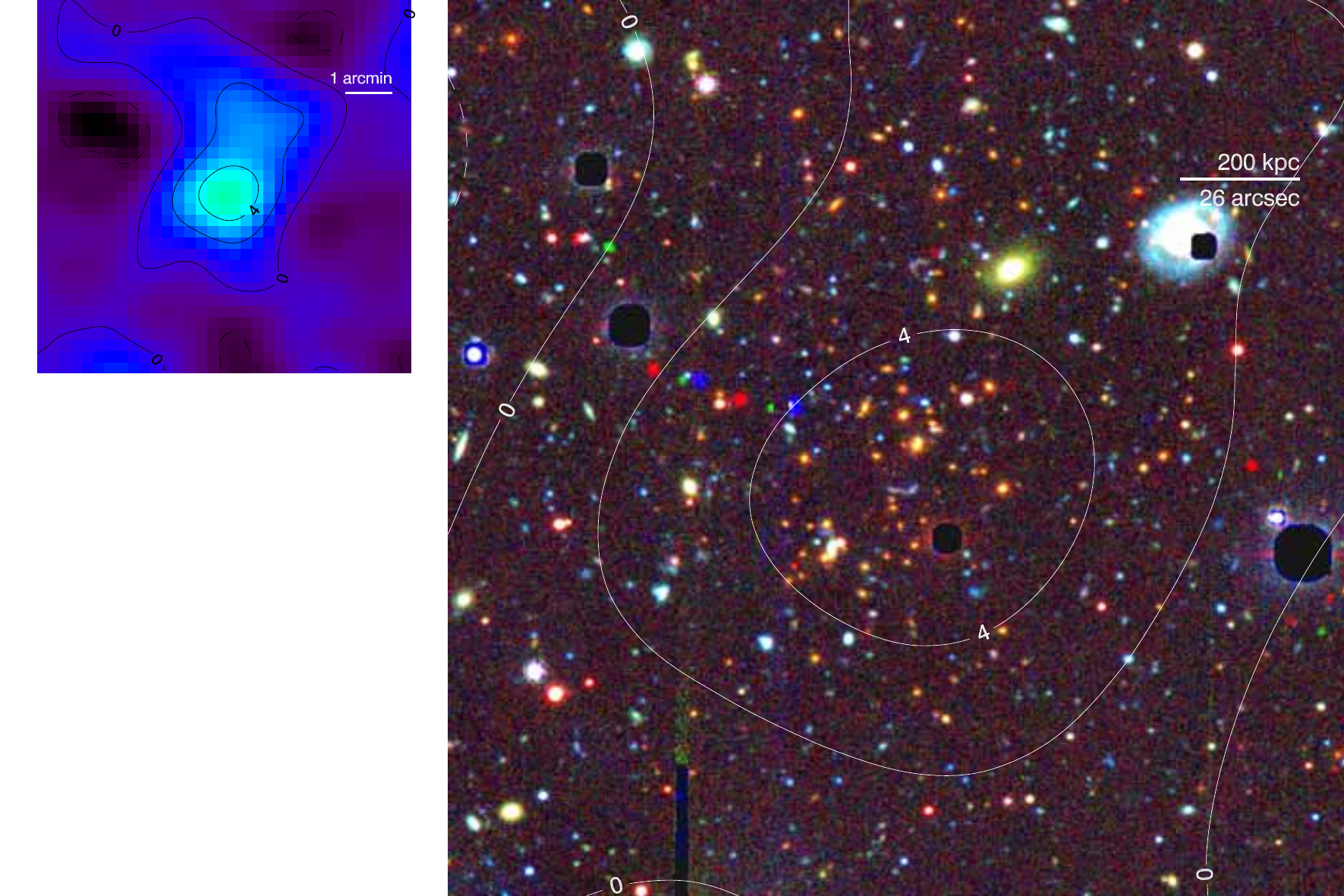}
 \caption{SPT-CL J0539-5744\label{fig:thumb7}}
\end{figure}

\begin{figure}
  \plotone{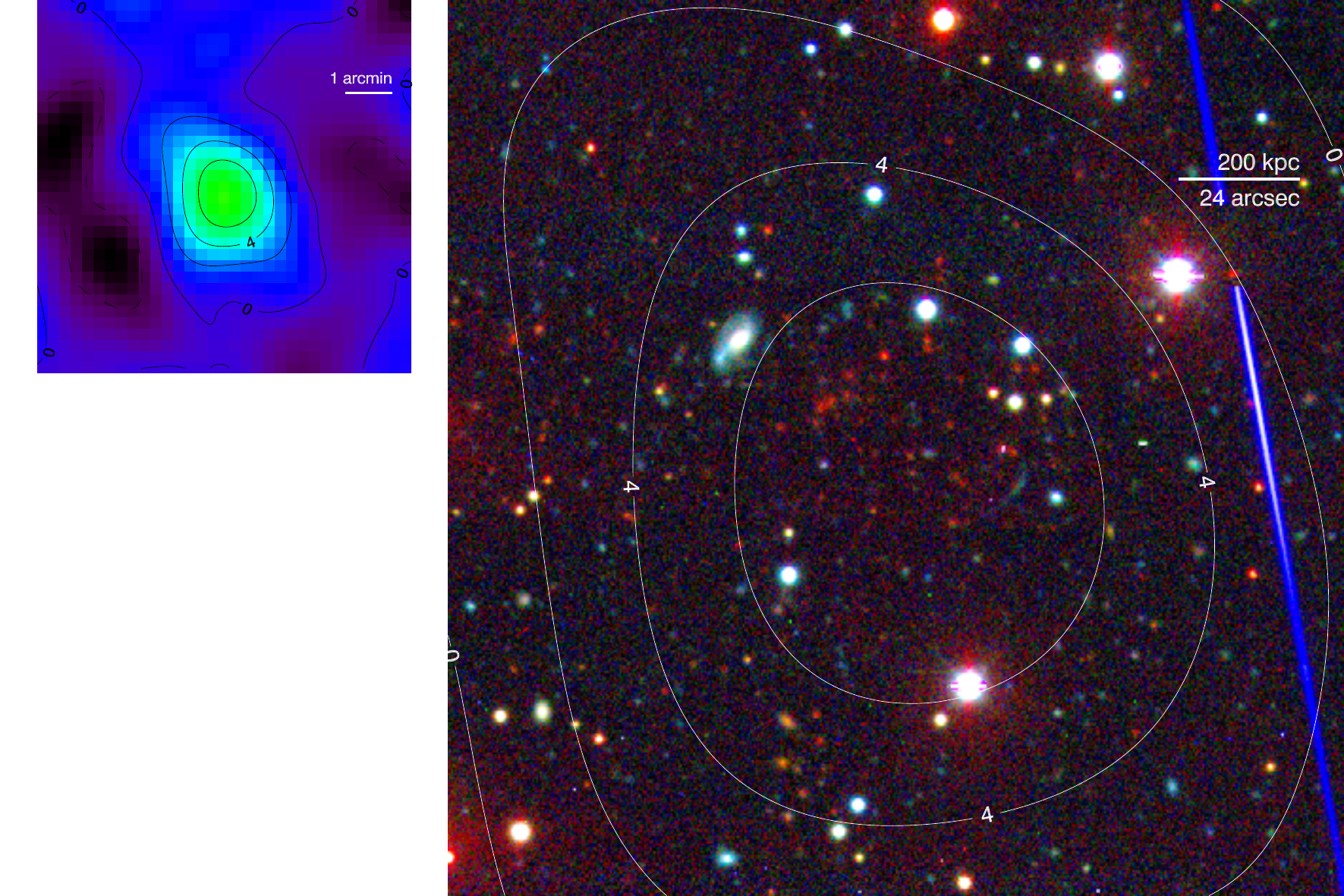}
 \caption{SPT-CL J0546-5345\label{fig:thumb8}}
\end{figure}

\begin{figure}
  \plotone{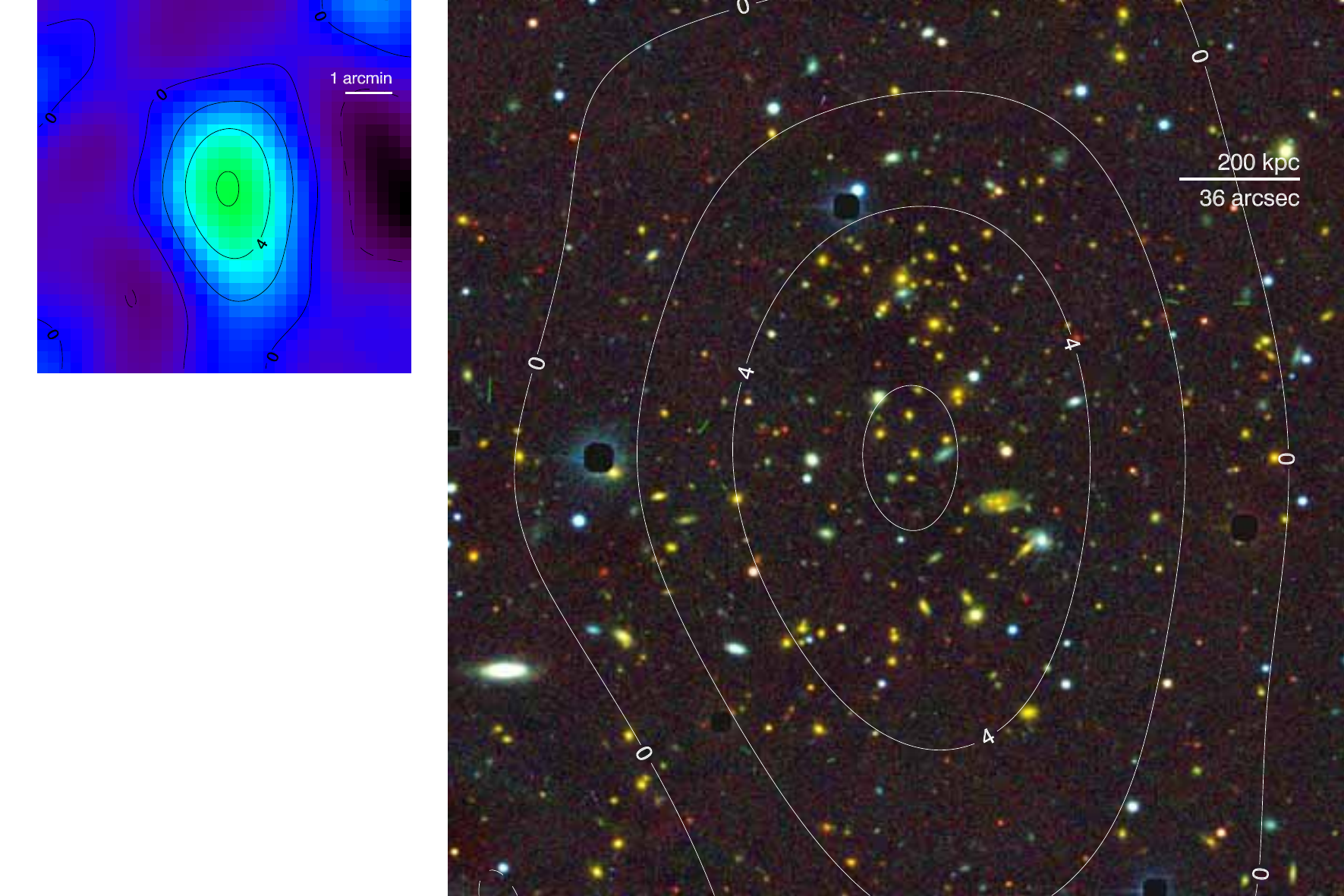}
 \caption{SPT-CL J0551-5709\label{fig:thumb9}}
\end{figure}

\begin{figure}
  \plotone{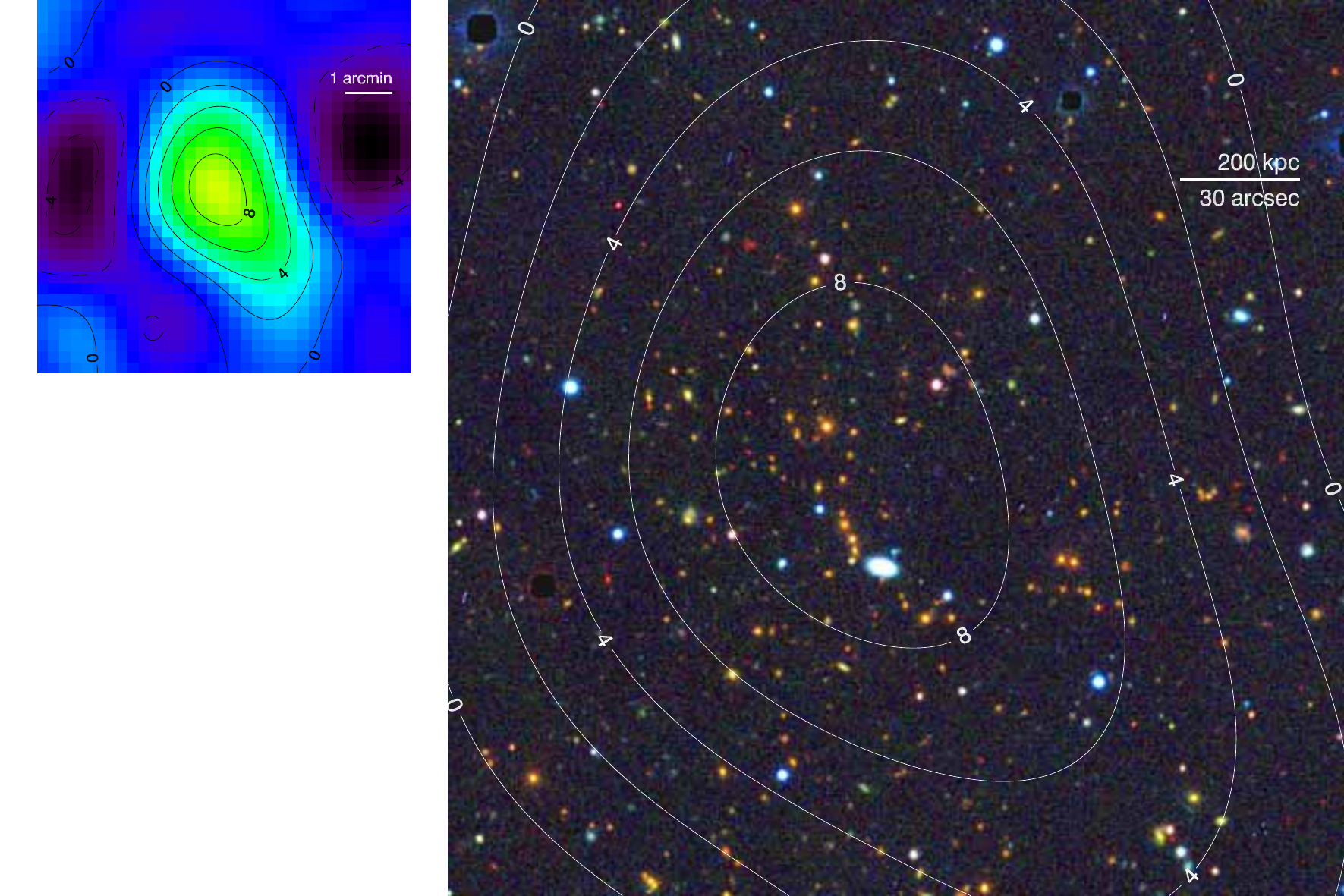}
 \caption{SPT-CL J0559-5249\label{fig:thumb10}}
\end{figure}

\begin{figure}
  \plotone{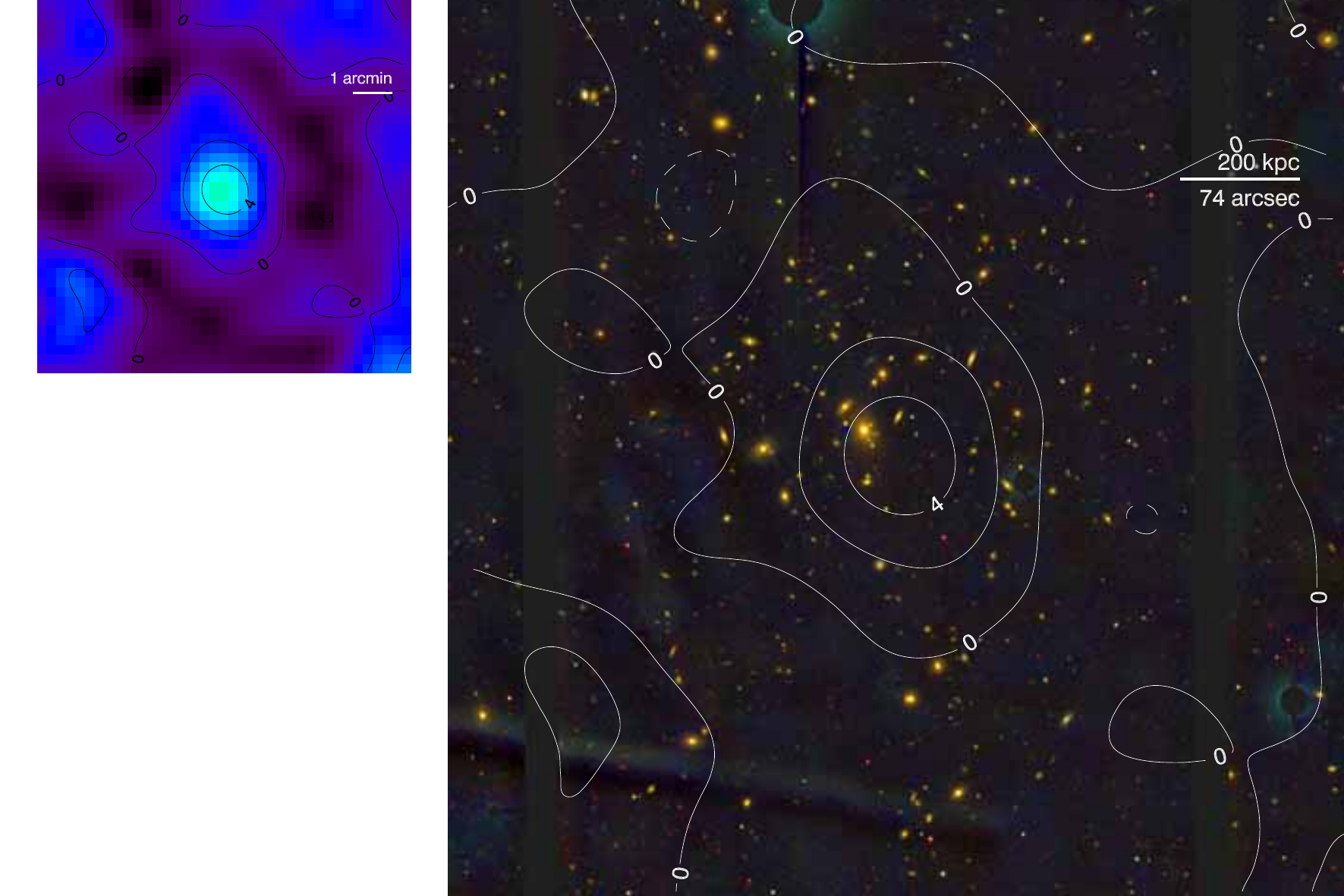}
 \caption{SPT-CL J2259-5617\label{fig:thumb11}}
\end{figure}

\begin{figure}
  \plotone{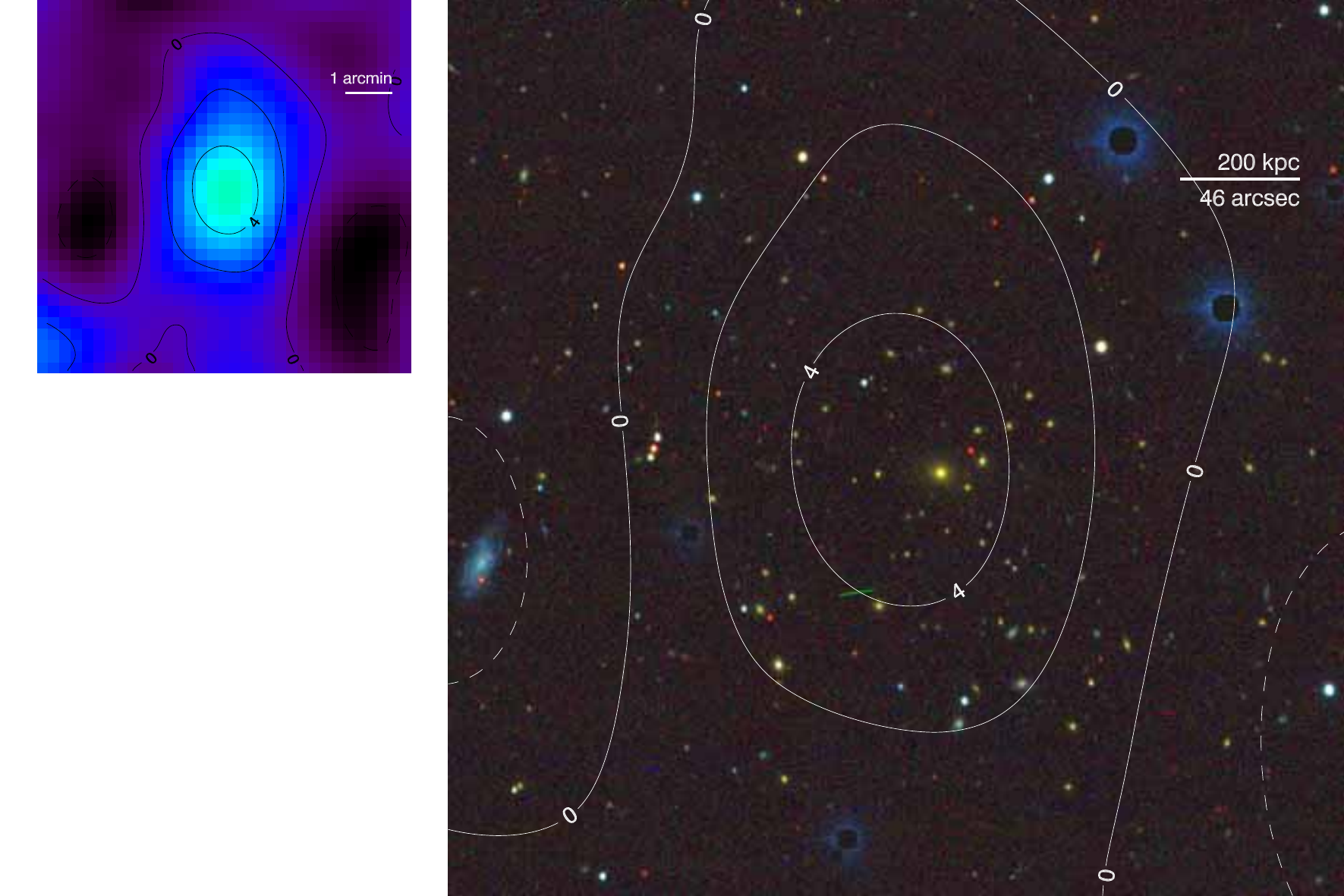}
 \caption{SPT-CL J2300-5331\label{fig:thumb12}}
\end{figure}

\begin{figure}
  \plotone{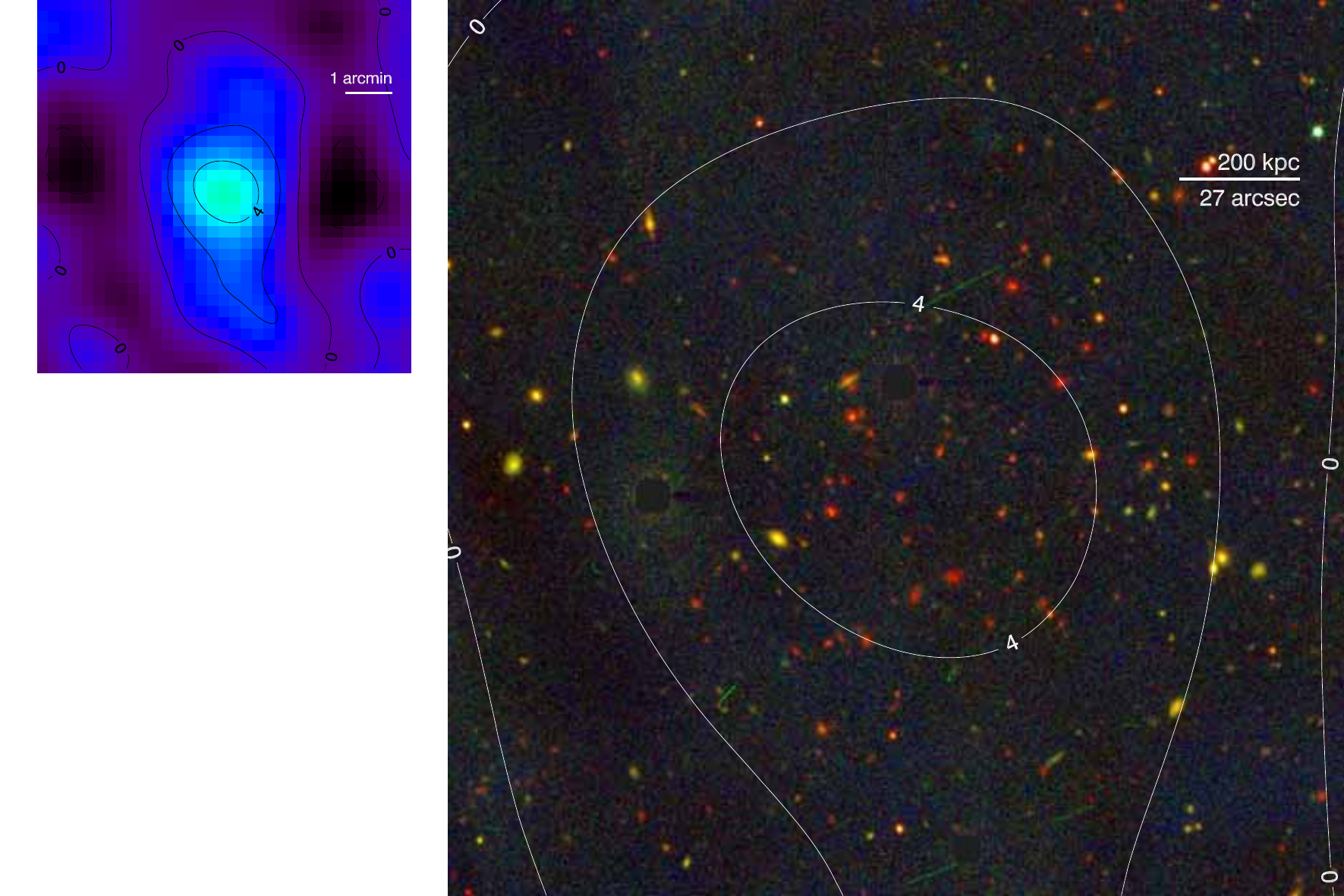}
 \caption{SPT-CL J2301-5546\label{fig:thumb13}}
\end{figure}

\begin{figure}
  \plotone{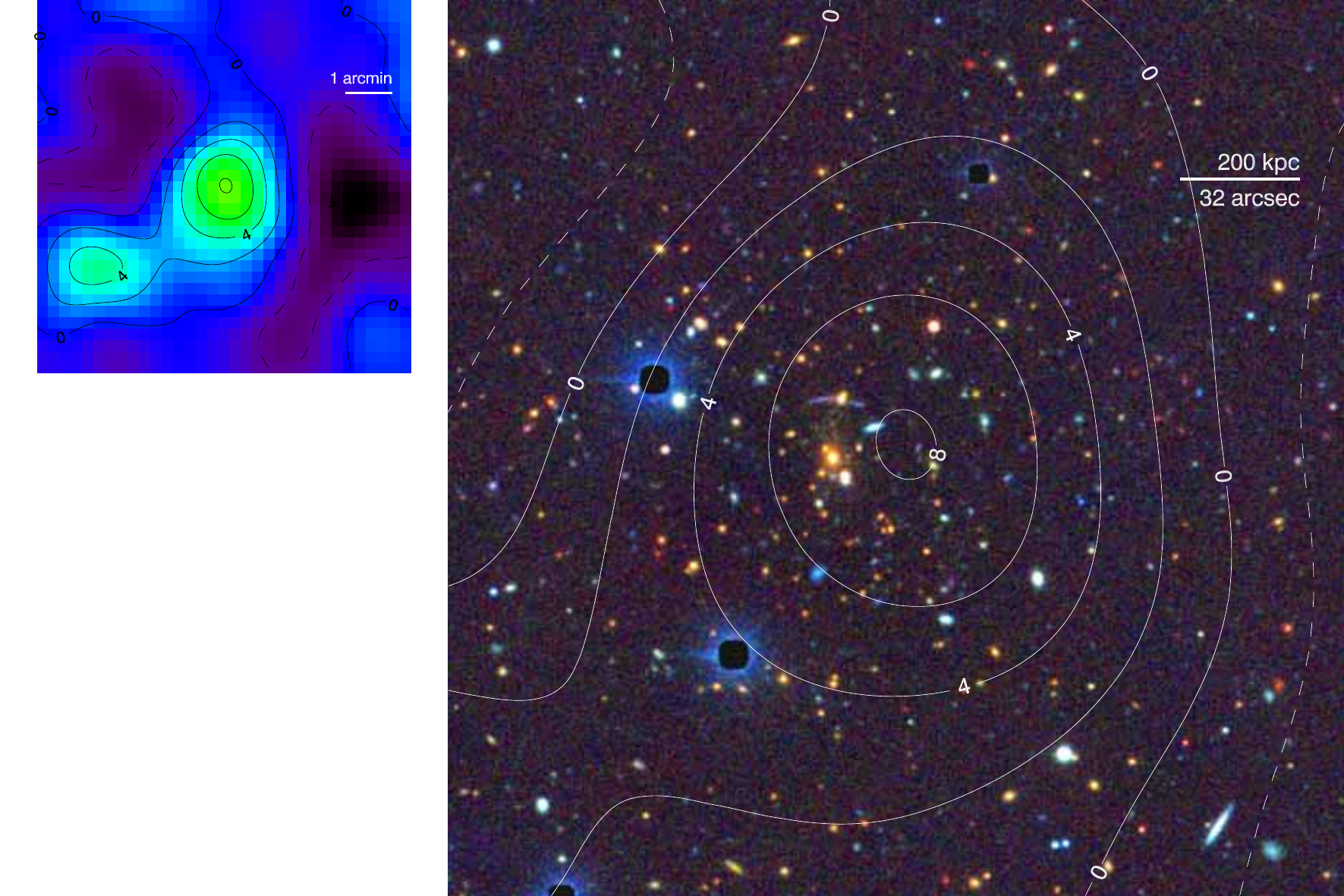}
 \caption{SPT-CL J2331-5051\label{fig:thumb14}}
\end{figure}

\begin{figure}
  \plotone{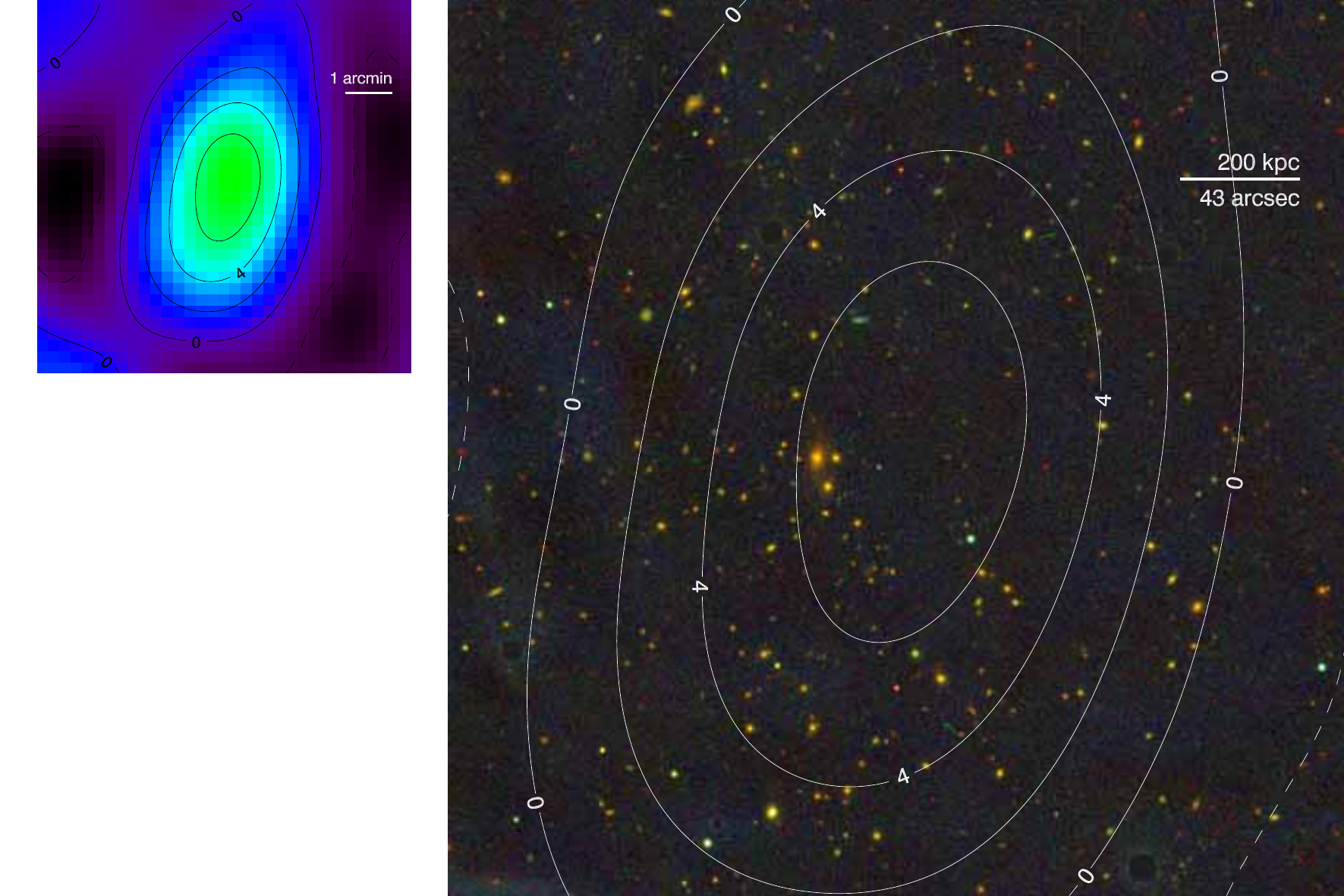}
 \caption{SPT-CL J2332-5358\label{fig:thumb15}}
\end{figure}

\begin{figure}
  \plotone{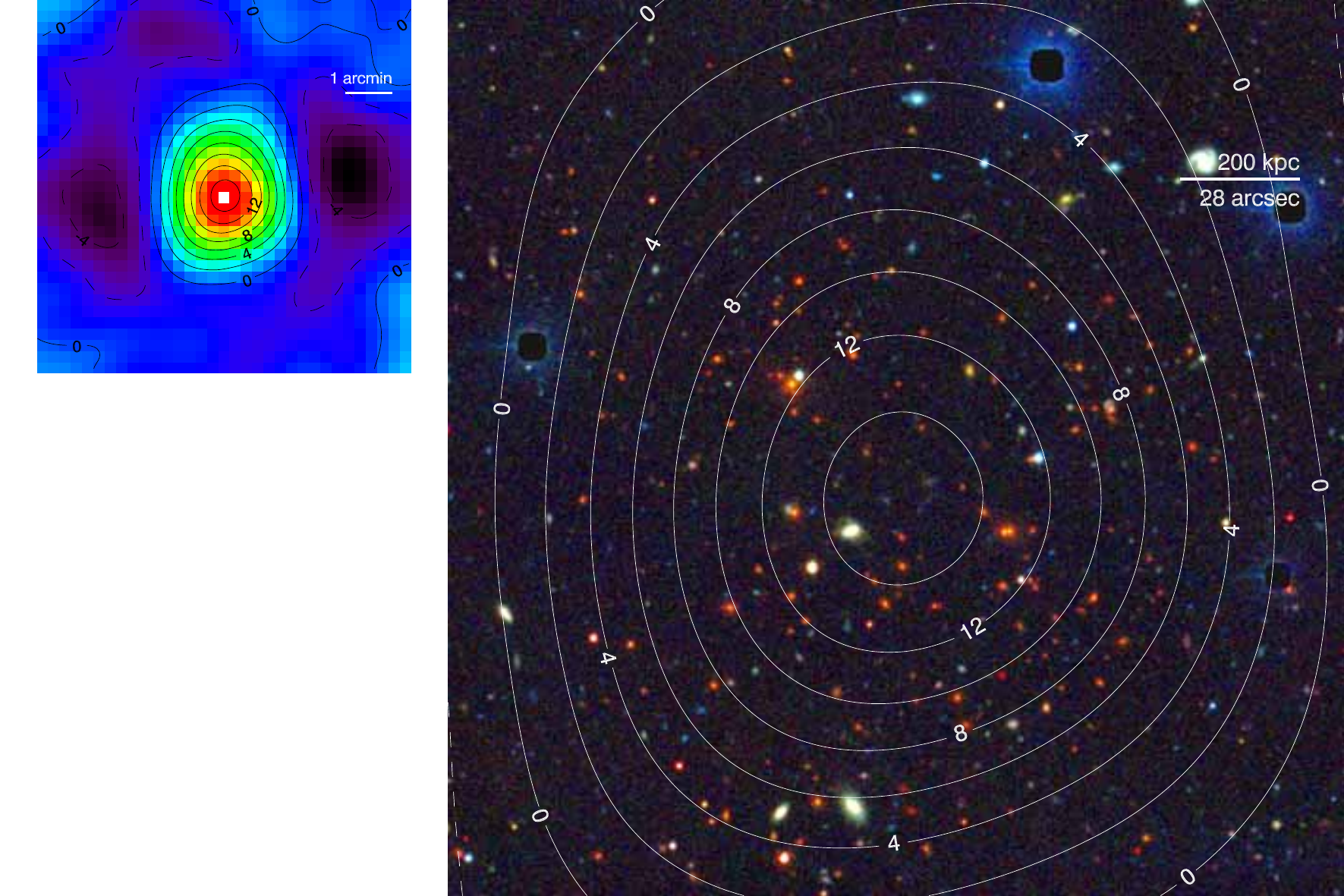}
 \caption{SPT-CL J2337-5942\label{fig:thumb16}}
\end{figure}

\begin{figure}
  \plotone{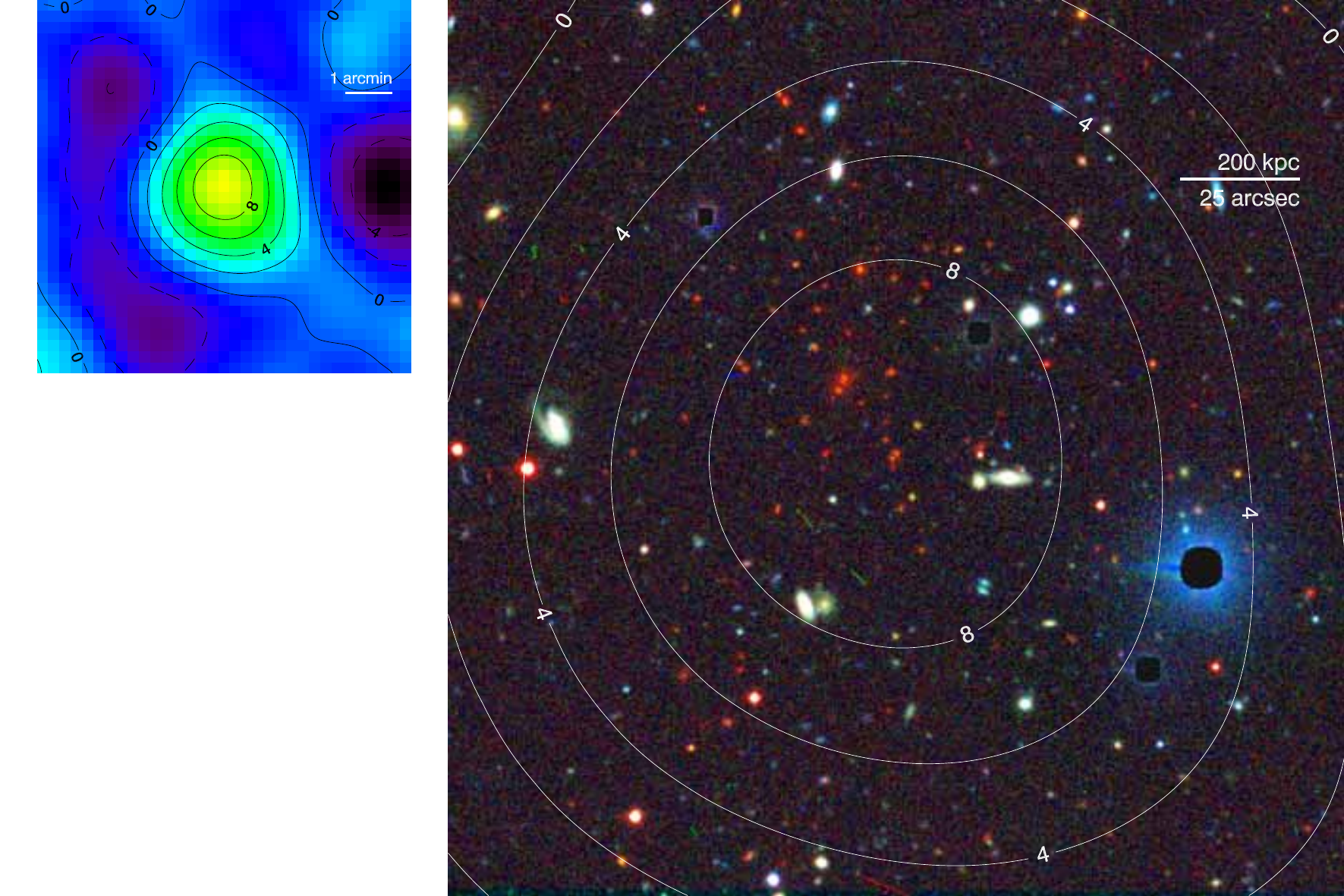}
 \caption{SPT-CL J2341-5119\label{fig:thumb17}}
\end{figure}

\begin{figure}
  \plotone{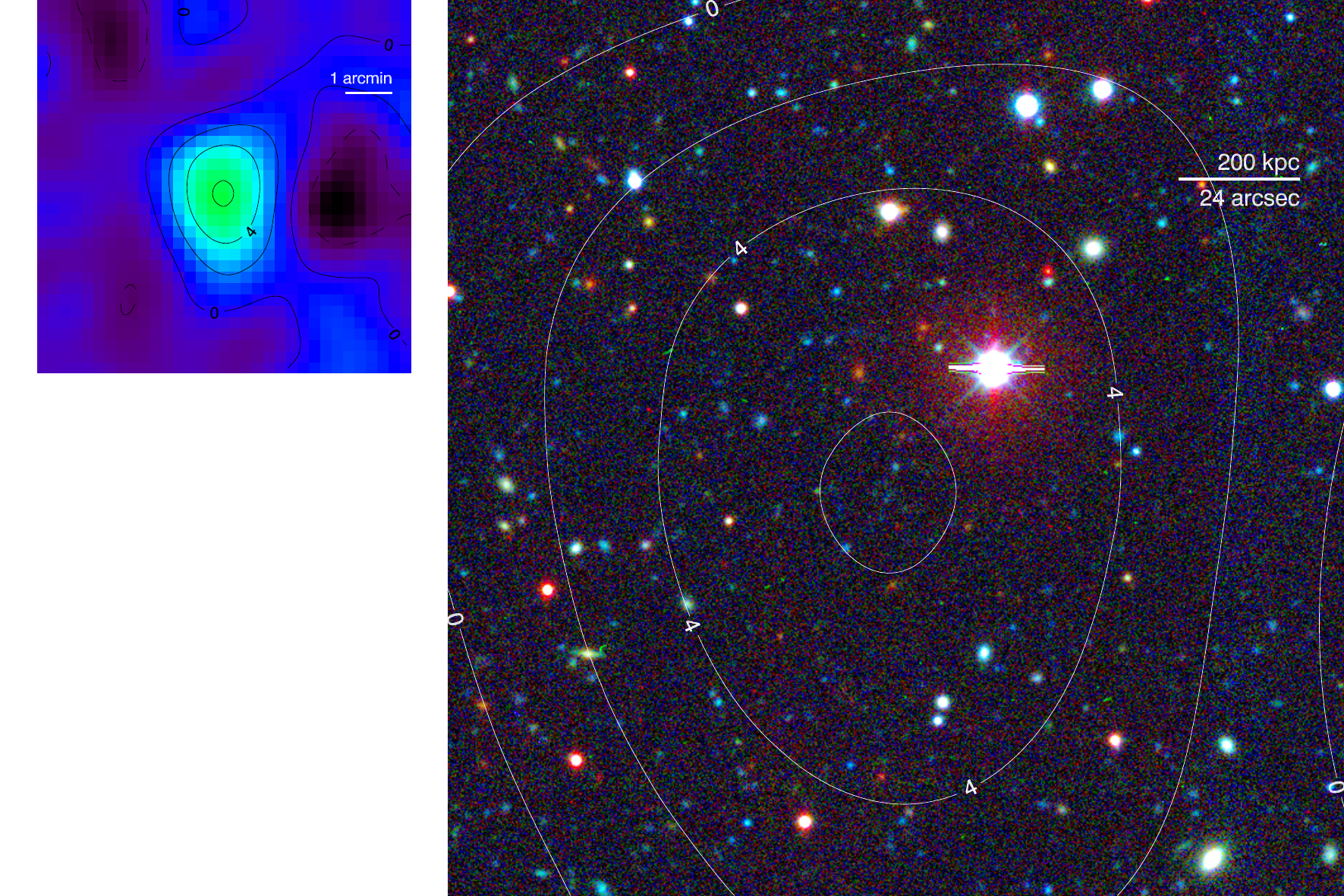}
 \caption{SPT-CL J2342-5411\label{fig:thumb18}}
\end{figure}

\begin{figure}
  \plotone{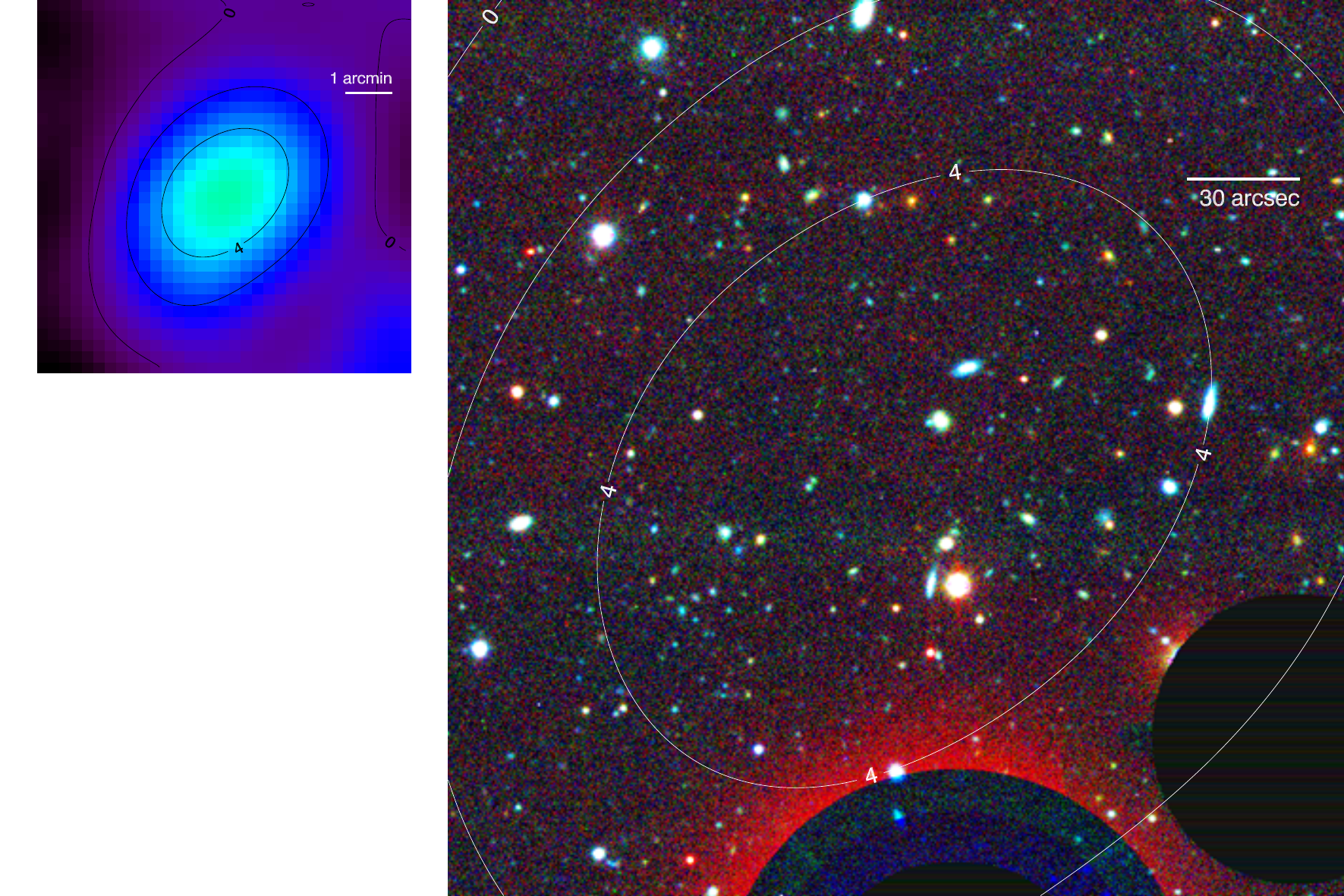}
 \caption{SPT-CL J2343-5521 (unconfirmed)\label{fig:thumb19}}
\end{figure}

\begin{figure}
  \plotone{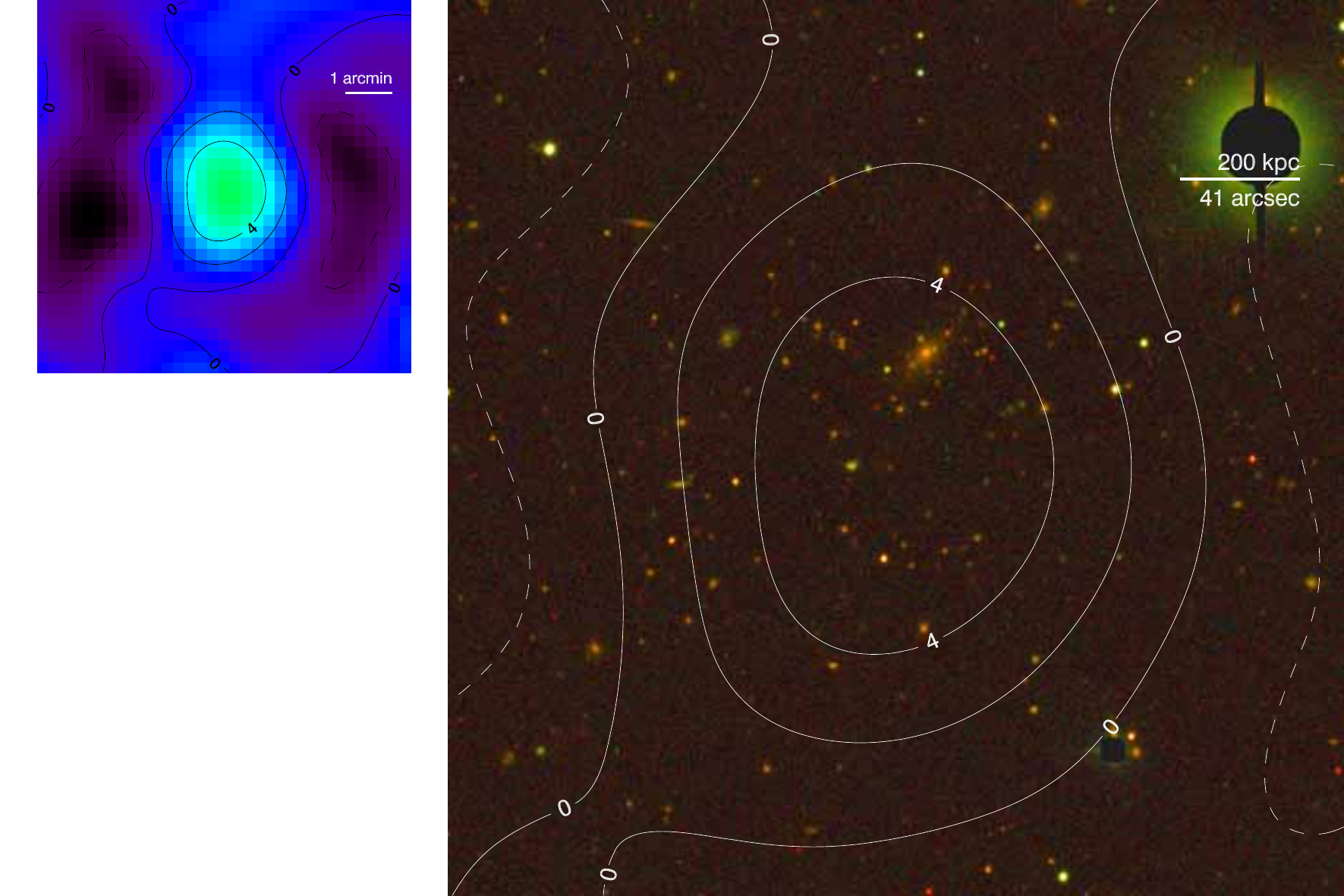}
 \caption{SPT-CL J2355-5056\label{fig:thumb20}}
\end{figure}

\begin{figure}
  \plotone{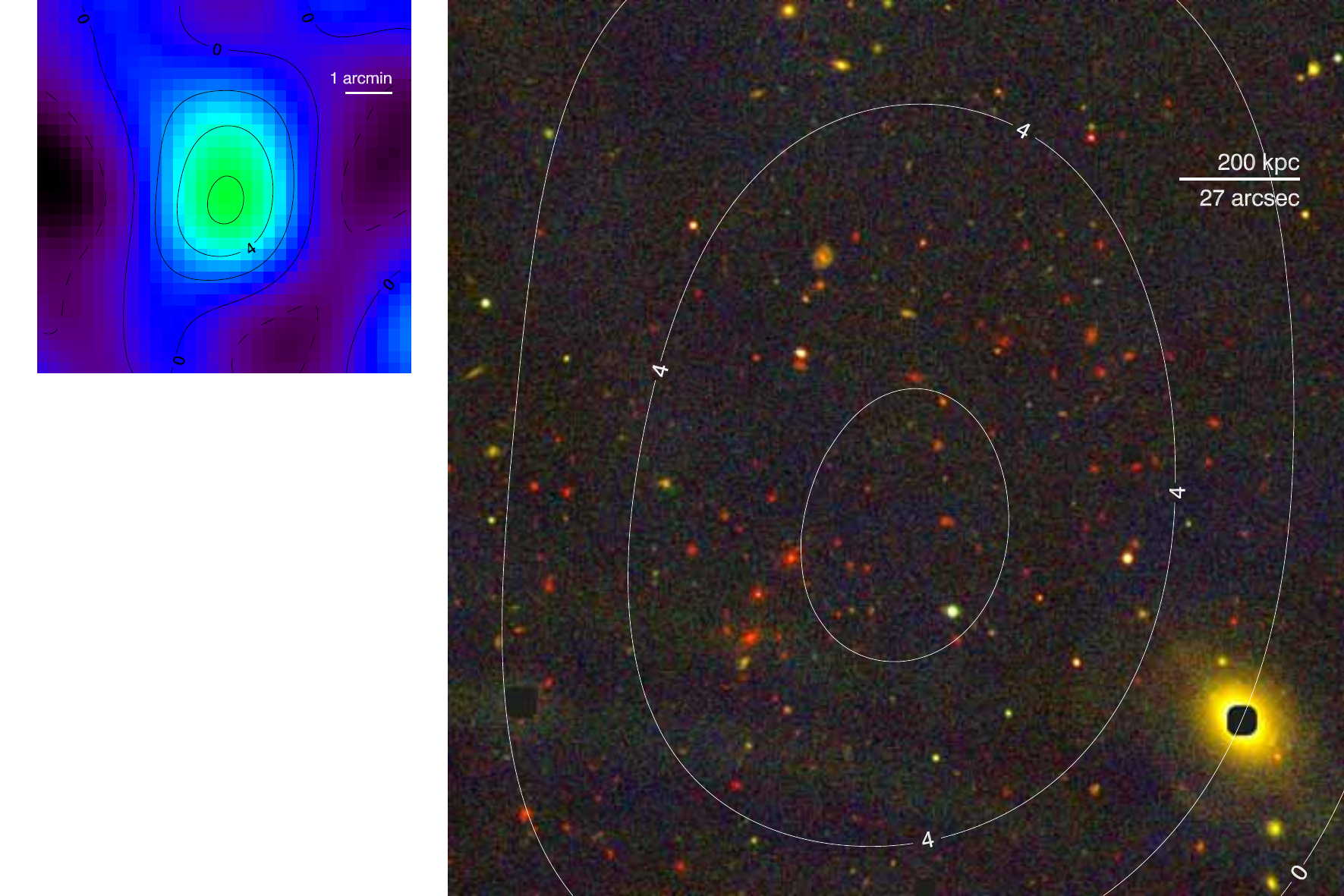}
 \caption{SPT-CL J2359-5009\label{fig:thumb21}}
\end{figure}

\begin{figure}
  \plotone{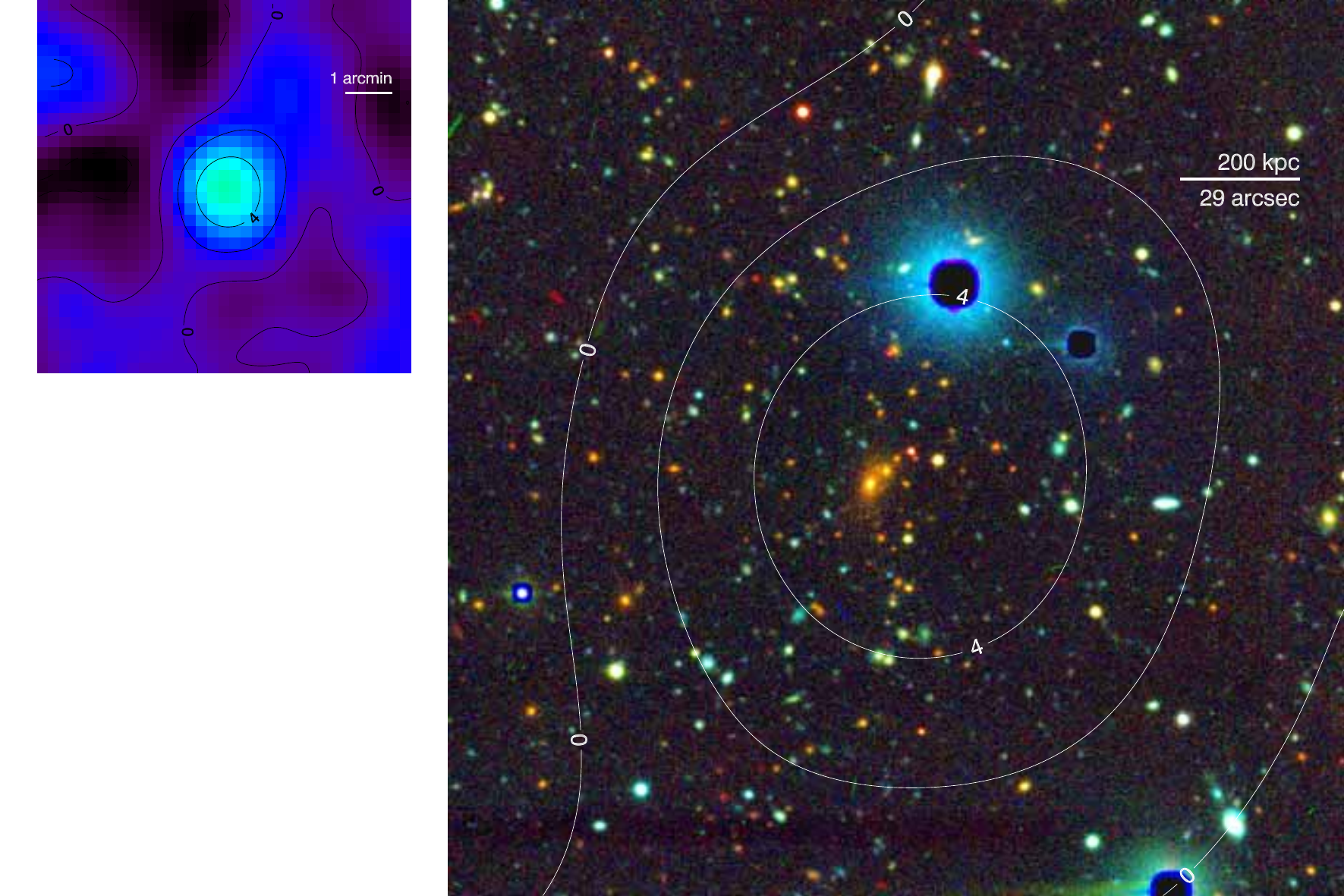}
 \caption{SPT-CL J0000-5748\label{fig:thumb22}}
\end{figure}

Figures \ref{fig:thumb1}--\ref{fig:thumb22} are false-color optical
images of the clusters, together with the SZ detection significance
maps.  In all images, north is up, east is left.

The SZ-only insets subtend 8 arcminutes on a side. The mapping between
color and SZ significance $\xi$ is the same in all SZ thumbnails.  The
highest significance cluster, Figure \ref{fig:thumb16}, spans the
largest color range, which is $\xi\in[-5.57,14.94]$.  The peak value
in each thumbnail is equal to the quoted SZ detection significance in
\citetalias{vanderlinde10}.  Contours are given at steps of
$\Delta(\xi)=2$, and the corresponding values are labeled.  Contours
are dashed where $\xi$ is negative, and solid where $\xi$ is
nonnegative.

The optical images have the same contours as their corresponding SZ
thumbnail overlain, and subtend 1.5~Mpc in the cluster rest frame. We
have mapped the $zrg$ passbands to the RGB channels, respectively, so
that the apparent redness of red-sequence cluster galaxies
approximately increases with redshift.

\clearpage

\clearpage

\clearpage

\clearpage

\clearpage

\clearpage

\clearpage

\clearpage

\clearpage

\clearpage

\clearpage

